\documentclass[aps,prx,preprint,onecolumn,citeautoscript,superscriptaddress,nofootinbib,eqsecnum]{revtex4}
\usepackage{amsmath}
\usepackage{amssymb}
\usepackage{float}
\usepackage[section]{placeins}
\usepackage{bbm}
\usepackage{bm}
\usepackage{comment}
\usepackage{graphicx}
\usepackage{physics}
\usepackage{color}
\usepackage[papersize={8.5in,11in}]{geometry}
\usepackage[colorlinks=true]{hyperref}
\hypersetup{
    bookmarks=true,         
    unicode=false,          
    pdftoolbar=true,        
    pdfmenubar=true,        
    pdffitwindow=false,     
    pdfstartview={FitH},    
    pdftitle={S},    
    pdfauthor={S. Sachdev},     
    pdfsubject={},   
    pdfcreator={},   
    pdfproducer={}, 
    pdfkeywords={} {} {}, 
    pdfnewwindow=true,      
    colorlinks=true,       
    linkcolor=magenta, 
    citecolor=blue,        
    filecolor=magenta,      
    urlcolor=blue           
}

\usepackage{amsfonts}
\usepackage{upgreek}
\usepackage{slashed}
\usepackage{latexsym}
\usepackage[export]{adjustbox}
\usepackage{dsfont}
\usepackage{subcaption}
\newcommand{\beq}{\begin{eqnarray}}
\newcommand{\eeq}{\end{eqnarray}}

\newcommand{\im}{{\rm Im}}
\newcommand{\bsp}{\begin{split}}
\newcommand{\esp}{\end{split}}

\newcommand{\be}{\begin{equation}}
\newcommand{\ee}{\end{equation}}

\graphicspath{{./images/}}
\def\bea{\begin{eqnarray}}
\def\eea{\end{eqnarray}}

\makeatletter

\def\env@sqcases{%
  \let\@ifnextchar\new@ifnextchar
  \left\lbrack
  \def\arraystretch{1.2}%
  \array{@{}l@{\quad}l@{}}%
}
\makeatother

\begin{document}


\title{Electronic spectra with paramagnon fractionalization\\ in the single band Hubbard model}
\author{Eric Mascot}
\affiliation{Department of Physics, University of Illinois at Chicago, Chicago, IL 60607, USA}
\author{Alexander Nikolaenko}
\affiliation{Department of Physics, Harvard University, Cambridge MA-02138, USA}
\author{Maria Tikhanovskaya}
\affiliation{Department of Physics, Harvard University, Cambridge MA-02138, USA}
\author{Ya-Hui Zhang}
\affiliation{Department of Physics, Harvard University, Cambridge MA-02138, USA}
\author{Dirk K. Morr}
\affiliation{Department of Physics, University of Illinois at Chicago, Chicago, IL 60607, USA}
\author{Subir Sachdev}
\affiliation{Department of Physics, Harvard University, Cambridge MA-02138, USA}
\affiliation{School of Natural Sciences, Institute for Advanced Study, Princeton, NJ-08540, USA}

\date{\today}

\begin{abstract}
We examine spectral properties of a recently proposed theory of the intermediate temperature pseudogap metal phase of the cuprates. We show that this theory can be obtained from the familiar paramagnon theory of nearly antiferromagnetic metals by fractionalizing the paramagnon into two `hidden' layers of $S=1/2$ spins. The first hidden layer of spins hybridizes with the electrons as in a Kondo lattice heavy Fermi liquid, while the second hidden layer of spins forms a spin liquid with fractionalized spinon excitations. We compute the imaginary part of the electronic self energy induced by the spinon excitations. The energy and momentum dependence of the photoemission spectrum across the Brillouin zone provides a good match to observations by He {\it et al.} in Bi2201 (Science {\bf 331}, 1579 (2011)) and by Chen {\it et al.} in Bi2212 (Science {\bf 366}, 1099 (2019)).
\end{abstract}

\pacs{Valid PACS appear here}
\maketitle{}


\section{Introduction}
\label{sec:intro}

The diverse phenomena associated with high temperature superconductivity in the cuprates present a long-standing theoretical challenge. These materials support several distinct phases in the doping ($p$) and temperature ($T$) phase diagram.
Our focus in this paper will be on the nature of the intermediate temperature metallic states at small $p$, above the superconducting temperature. Such temperatures are well below the microscopic energy scales, and we maintain it is reasonable to describe them as quantum phases in their own right, instead of considering them as thermal fluctuations of the observed low temperature states. We will not consider here the low temperature plethora of charge and spin ordered states that have been explored in the underdoped cuprates. 

The large $p$ `overdoped' side is a system well described as a conventional Fermi liquid. In the small $p$ `underdoped' side, an enigmatic pseudogap phase arises at intermediate $T$. These phases have been thoroughly studied experimentally \cite{Proust2018}, including angle-resolved photoemission (ARPES) \cite{He2011,chen2019incoherent}, scanning tunneling microscopy (STM) \cite{Fischer2007}, transport \cite{Doiron-Leyraud2017} and thermodynamic measurements \cite{Michon2019}, but a complete theoretical understanding is still lacking. 

Many theoretical models have been proposed to  describe the pseudogap regime of the cuprate superconductors. Some of them assume that the pseudogap is a precursor to one or more ordered phases, such as a spin density wave (SDW) \cite{Chubukov1997}, or a charge density wave (CDW) \cite{KivelsonRMP}, or a pair density wave (PDW) \cite{Lee2014,AgterbergARCMP}, or a d-density wave (DDW) \cite{Chakravarty2001}. A different class of models assume that the pseudogap is a distinct phase of matter characterized by spin liquid physics, which likely undergoes a confinement crossover to a more conventional broken symmetry at low temperatures. 

Here, we shall investigate a model in the latter class: the model describes the pseudogap metal as a state which has electron-like quasiparticles around a Luttinger-rule violating small Fermi surface, along with neutral spinon excitations (such states are sometimes called fractionalized Fermi liquids (FL*)). We will show that a recently introduced theory of such a phase in a single band Hubbard model \cite{Zhang2020,Zhang2020_1,nikolaenko2021} yields simple models which can be successfully compared to a wide range of ARPES observations in Bi2212 and Bi2201, in both the nodal and anti-nodal regions of the Brillouin zone \cite{chen2019incoherent,He2011}. We reiterate that we do not expect the fractionalization to survive down to zero temperature, and there is likely a confinement crossover/transition as the temperature is lowered, such as those discussed in Refs.~\cite{Patel:2016efz,SCSS16,Christos:2021wno}

We will begin in Section~\ref{sec:ancilla} by recalling the framework of the theory in a manner which displays its close connection to the familiar paramagnon theory of magnetism in metals, and show that it is a theory of paramagnon fractionalization.
Section~\ref{sec:frac} and Appendix~\ref{app:z2} present results on the self-energy of the electrons in the pseudogap metal from their coupling to the second hidden layer of spins produced by paramagnon fractionalization. 
Our theory will be applied to ARPES observations on the single-layer cuprate compound Bi2201 by He {\it et al.} \cite{He2011} in Section~\ref{sec:2201}; we will also mention connections to STM observations in Section~\ref{sec:stm}. Then 
we will turn to recent observations on the bilayer cuprate compound Bi2212 by Chen {\it et al.} \cite{chen2019incoherent} in Section~\ref{sec:2212}.

\section{Paramagnon fractionalization}
\label{sec:ancilla}

We begin by presenting a review of the main ideas of the theory employing `ancilla' or `hidden' qubits \cite{Zhang2020,Zhang2020_1,nikolaenko2021}, highlighting here its close connection to the popular paramagnon theory of magnetic order in Fermi liquids \cite{Berk66,Doniach66,Scalapino_dwave,ChubukovAP}. The starting assertion of this theory is that we should {\it not\/} use the common ansatz for the fractionalization of the electron $c_\alpha$ into a spinon $f_\alpha$ and holon $b$, $c_\alpha = f_\alpha b^\dagger$. We avoid this particular fractionalization because the excitations around the small Fermi surface carry spin-1/2 and charge $e$, just like the bare electron. 
So it is cumbersome to fractionalize all the electrons, and then undo the fractionalization for a small density of them by forming bound states of spinons and holons. 
In a $t$-$J$ model (with hopping $t$ and exchange $J$), 
the charged excitations could remain fractionalized at the lowest energies for $J \gg t$, but for $t \gg J$ the hopping provides a strong attractive potential for binding spinons and holons.
Such a spinon-holon binding route has been proposed in many models of the pseudogap metal, see {\em e.g.\/}
Refs.~\cite{WenLee96,Kaul08,Scheurer:2017jcp}.
But no theory of this bound state formation at non-zero hole density has yet been presented, although there have been some notable works \cite{Ribeiro05,Punk:2015fha,Bohrdt:2020zsd,Bohrdt:2021axw}.

We describe the structure of the theory here by writing the single-band Hubbard model in terms of nearly-free electrons coupled to paramagnons. Indeed, this is the starting point of numerous theories for SDW order, and also for fluctuating SDW order in metals. The main new idea of our theory is to fractionalize the paramagnon. 

First, we recall the paramagnon theory of magnetism in metals, and cast it in Hamiltonian form. We begin with the Hubbard model 
\beq
H_U = - \sum_{i<j} t_{ij} \left[ c_{i \alpha}^\dagger c_{j \alpha} + c_{j \alpha}^\dagger c_{i \alpha} \right] + \sum_i \left[ - \mu (
n_{i \uparrow} + n_{i \downarrow}) + U 
n_{i \uparrow} n_{i \downarrow}\right]\label{dwave1}
\eeq
where $c_{i \alpha}$ annihilates a fermion on site $i$ with spin $\alpha = \uparrow, \downarrow$, and
\beq
n_{i \uparrow} = c_{i \uparrow}^\dagger c_{i \uparrow} \quad, \quad n_{i \downarrow} = c_{i \downarrow}^\dagger c_{i \downarrow}. 
\eeq
After using the single-site identity
\begin{equation}
U \left(n_{i\uparrow} - \frac{1}{2} \right) \left(n_{i\downarrow} - \frac{1}{2} \right) = -\frac{2U}{3} {\bm S}_i^{2} + \frac{U}{4} \,,
\end{equation}
(which is easily established from the electron commutation relations) it becomes possible to decouple the 4-fermion term in a particle-hole channel. We perform the Hubbard-Stratonovich transformation
\begin{equation} 
\exp \left( \frac{2U}{3} \sum_i \int d \tau {\bm S}^{2}_i \right) = \int \mathcal{D} {\bm \Phi}_i (\tau) \exp
\left( - \sum_i \int d \tau \left[ \frac{3}{8U} {\bm \Phi}^{2}_i + {\bm \Phi}_i \cdot c_{i \alpha}^\dagger \frac{{\bm \sigma}_{\alpha\beta}}{2} c_{i \beta} \right] \right) \,,
\end{equation}
and obtain a new field ${\bm \Phi}_i (\tau)$ which will play the role of the paramagnon. 

It is useful to write down the paramagnon theory in Hamiltonian form, without integrating out the low energy electron modes. To make the ${\Phi}_i$ field dynamical, we can integrate some of the {\it high\/} energy electrons far from the Fermi surface. This will 
induce additional terms in a local potential $V({\bm \Phi}_i)$ which controls fluctuations in the magnitude of ${\bm \Phi}_i$. The high energy electrons will also induce a dynamical term $\omega_n^2 {\bm \Phi}_i^2$ in the effective action for ${\bm \Phi}_i$ (where $\omega_n$ is a Matsubara frequency) so that the on-site ${\bm \Phi}_i$ paramagnon Lagrangian on each site is
\beq
\mathcal{L}_i = \frac{1}{2g} \left( \partial_\tau {\bm \Phi}_i \right)^2 + V ( {\bm \Phi}_i) \label{LPhi}
\eeq
Our main assumption is that $V({\bm \Phi}_i)$ has a
minimum at a non-zero value of $|{\bm \Phi}_i|$ so that the low energy states correspond to the angular momentum $\ell = 0, 1 \ldots$ states from rotational motion of the ${\bm \Phi}_i$ in a radial state around the $|{\bm \Phi}_i|$ minimum. In the following, we will only keep the $\ell =0,1$ states.

\begin{figure}
\begin{center}
\includegraphics[width=3.5in]{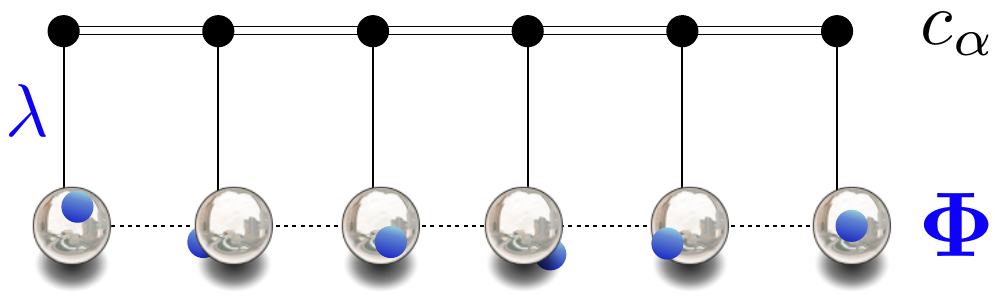}
\end{center}
\caption{Hamiltonian form of the paramagnon theory of magnetism in Fermi liquids: a band of electrons $c_{\alpha}$, with each site coupled to a paramagnon quantum rotor. The rotor is a particle of mass $1/g$ constrained to move on the unit sphere with co-ordinate ${\bm \Phi}$. The $c_\alpha$ and ${\bm \Phi}$ reside on a $d>1$ dimensional lattice, although only one dimension is shown.}
\label{fig:rotor}
\end{figure}

We can make the analysis more explicit by rescaling $|{\bm \Phi}_i|$, and 
replacing $V({\bm \Phi}_i)$ by a unit-length constraint on each site
\beq
{\bm \Phi}_i^2 = 1\,; \label{Phiconstraint}
\eeq
we emphasize that this constraint is not essential to any of the following analysis, and merely a simplifying choice.
With the constraint 
Eq.~(\ref{Phiconstraint}), the dynamical term in Eq.~(\ref{LPhi}) is simply the kinetic energy $\propto {\bm L}_i^2$ of a particle with angular momentum ${\bm L}_i$ moving on the unit sphere. 
In this manner, we obtain a Hamiltonian for electrons coupled to a paramagnon {\it quantum rotor\/} on each site, illustrated in Fig.~\ref{fig:rotor}.
\beq
H_{\rm paramagnon} = \sum_{{\bm p}}  \varepsilon_{\bm p} c_{{\bm p}\sigma} ^\dagger c_{{\bm p}\sigma} 
+  \frac{g}{2} \sum_{i} {\bm L}_i^2 
+ \sum_i \left( \lambda {\bm \Phi}_{i} + \widetilde{\lambda} {\bm L}_i \right)  \cdot c_{i \alpha}^\dagger \frac{{\bm \sigma}_{\alpha\beta}}{2} c_{i \beta}\,. \label{Hrotor}
\eeq
The ${\bm L}_i$ and ${\bm \Phi}_i$ obey the usual commutation relations of single-particle rotational quantum mechanics
\beq
[L_a, L_b] = i \epsilon_{abc} L_c \quad, \quad [L_a, \Phi_b] = i \epsilon_{abc} \Phi_c \quad, \quad [\Phi_a, \Phi_b] = 0\,,
\eeq
where $a,b,c=x,y,z$, we have dropped the site label $i$, and $\epsilon_{abc}$ is the unit antisymmetric tensor.
Now on each site we have a `particle' of mass $1/g$ moving on a unit sphere with angular momentum ${\bm L}_i$; this is the paramagnon rotor, which has couplings $\lambda$, $\widetilde{\lambda}$ to the low energy electrons. 

To ensure the consistency of our procedure, let us undo the mappings above, and show how the original Hubbard model can be obtained starting from the rotor-fermion Hamiltonian in Eq.~(\ref{Hrotor}). At $\lambda, \widetilde{\lambda}=0$, it is a simple matter to diagonalize the rotor spectrum. On each site, we have states labelled by the usual angular momentum quantum numbers $\ell$
\beq
\left| \ell, m \right\rangle_i \quad, \quad \ell=0,1,2,\ldots; m = - \ell, \ldots \ell \quad , \quad \mbox{Energy} = \frac{g}{2} \ell (\ell + 1)\,.
\label{rotorl}
\eeq
So the ground state has $\ell = 0$ on each site, and there is a 3-fold degenerate excited state with $\ell =1$ and energy $g$. 
Turning to non-zero $\lambda$, but $ |\lambda| \ll g$, we can eliminate the coupling between the electrons and the rotors by a canonical transformation. Note that the $\lambda$ coupling is only active when the electronic state on a given site $i$ has spin 1/2: so the influence of the $\lambda$ is to lower the energy of this state with respect to the empty and doubly occupied sites which has spin 0. To compute this energy shift we need the matrix elements of ${\bm \Phi}$ between the $\ell =0$ ground state and the $\ell=1$ excited states; the non-zero matrix elements are
\beq
\left\langle \ell =0 \right| \Phi_z \left| \ell =1, m=0 \right\rangle = \frac{1}{\sqrt{3}}  , \quad \left\langle \ell =0 \right| \Phi_x \pm i \Phi_y \left| \ell =1, m=\mp 1 \right\rangle = \sqrt{\frac{2}{3}}\,.
\label{rotormat}
\eeq
We can now use perturbation theory in $\lambda$ to compute the energy of the spin 1/2 state; in this manner,
to second-order in $\lambda$, we obtain an effective model which is just the original Hubbard model with 
\beq
U = \frac{ \lambda^2}{4g}\,.
\eeq
Note that the coupling $\widetilde{\lambda}$ does not appear at this order, because ${\bm L}_i = 0$ when acting on the rung singlet state.

It is important to note that, so far, all we have done is to cast the familiar paramagnon theory in a Hamiltonian form. No fundamentally new step has yet been taken.

\begin{figure}
\includegraphics[width=5in]{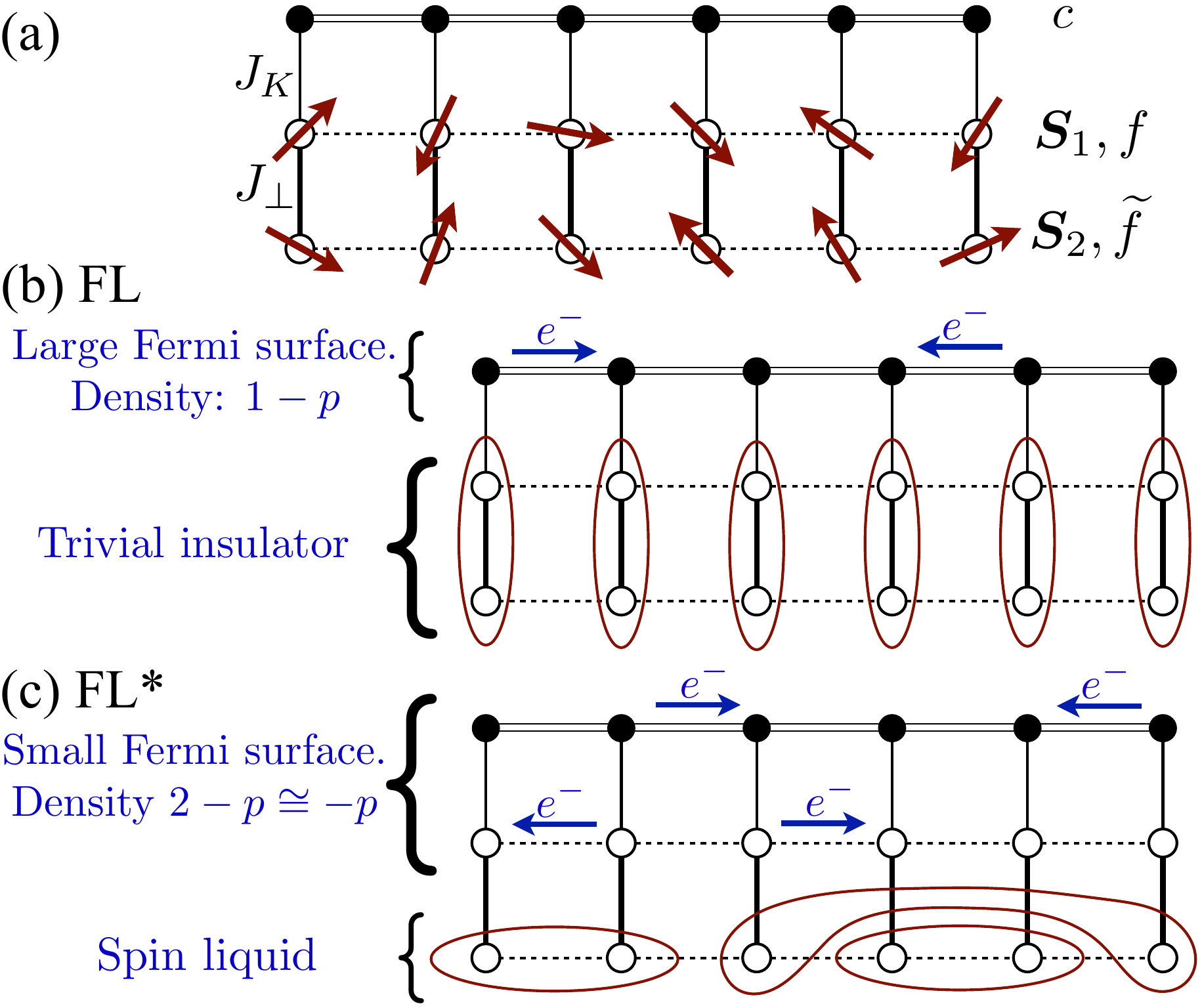}
\caption{We realize the paramagnon rotors in Fig.~\ref{fig:rotor} by a pair of qubits (spin-1/2 spins) represented by Schwinger fermions $f$ and $\widetilde{f}$. We indicate the antiferromagnetic exchange couplings $J_K$ and $J_\perp$, and the dashed lines represent exchange interactions within the $f$ and $\widetilde{f}$ layers. There is also an exchange interaction $\widetilde{J}_K$ in Eq.~(\ref{tJKval}) between the $c$ and $\widetilde{f}$ fermions which is not shown. The $\widetilde{f}$ spins form a spin liquid with fractionalization in the FL* phase, whose presence is required by the generalized Luttinger relation. }
\label{fig:layers}
\end{figure}

The new step is to replace each paramagnon rotor by a pair of antiferromagnetically coupled spins. We restrict attention to only the lowest energy $\ell = 0,1$ angular momentum states of each rotor in Eq.~(\ref{rotorl}). We can represent these singlet and triplet states by a pair of $S=1/2$ spins, ${\bm S}_{1i}$, ${\bm S}_{2i}$ coupled with an antiferromagnetic exchange coupling 
\beq
J_\perp = g \,. 
\eeq
This yields the singlet and triplet states with the energy splitting in Eq.~(\ref{rotorl}), and the operator correspondence which reproduces the matrix elements in Eq.~(\ref{rotormat}) in this subspace is
\bea
{\bm L}_i &=& {\bm S}_{1i} + {\bm S}_{2i} \nonumber \\
{\bm \Phi}_i &=& \frac{1}{\sqrt{3}} \left({\bm S}_{1i} - {\bm S}_{2i} \right)\,.
\eea
These spins are the `ancilla' or `hidden' qubits, as illustrated in Fig.~\ref{fig:layers} in which the top physical layer of electrons, $c$, of density $1-p$ is coupled to 2 layers of qubits, replacing the paramagnon rotors in Fig.~\ref{fig:rotor}.
The Kondo coupling $J_K$ between the $c$ spin and ${\bm S}_1$ layers in Fig.~\ref{fig:layers} is 
\be
J_K = \lambda/\sqrt{3} + \widetilde{\lambda}\,. \label{JKval}
\ee
Similarly there is a coupling 
\beq
\widetilde{J}_K = -\lambda/\sqrt{3} + \widetilde{\lambda} \label{tJKval}
\eeq
between the $c$ spin and ${\bm S}_2$, but we don't display it for simplicity.

Finally, we use the Schwinger fermion representation 
for both layers of hidden spins
\beq
{\bm S}_{1i} = \frac{1}{2} f_{i; \alpha}^\dagger {\bm \sigma}_{\alpha\beta} f_{ i; \beta}\, \quad, \quad {\bm S}_{2i} = \frac{1}{2} \widetilde{f}_{i; \alpha}^\dagger {\bm \sigma}_{\alpha\beta} \widetilde{f}_{ i; \beta}\,,
\eeq
with the constraints
\beq
\sum_{\alpha} f_{i;\alpha}^\dagger f_{i;\alpha} = 1\,\quad, \quad \sum_{\alpha} \widetilde{f}_{i;\alpha}^\dagger \widetilde{f}_{i;\alpha} = 1 \label{Psiunit}
\eeq
satisfied on each lattice site $i$. 
It is important that we add {\it two} layers of qubits, because only then are the added layers allowed to form a trivial insulator. 

This discussion should make it clear that we have not add any new low energy degrees of freedom, just as was the case in the paramagnon theory. Indeed, because we truncated Hilbert space of each rotor to the $\ell = 0,1$ states, the present theory actually contains fewer states in the full Hilbert space than the paramagnon theory. 

In the large Fermi surface FL phase in Fig.~\ref{fig:layers}a, we assume that the antiferromagnetic coupling $J_\perp$ dominates, and so the hidden qubits are locked into rung singlets, and can be safely ignored in the low energy theory: then the $c$ electrons form a conventional Fermi liquid phase \cite{vanilla}, and we obtain a Fermi surface corresponding to electron density $1-p$, or hole density $1+p$. The singlet to triplet excitation of the rung spins is precisely the paramagnon ${\bm \Phi}$ particle, and so the theory reduces here to the familiar paramagnon theory of correlated metals \cite{Scalapino_dwave,ChubukovAP}.  

In the small Fermi surface FL* phase in  Fig.~\ref{fig:layers}b, we assume that the antiferromagnetic Kondo coupling $J_K$ dominates, and so the Kondo effect causes the $f$ spins to `dissolve' into the Fermi sea of the mobile electrons. By analogy with the heavy Fermi liquid phase of the Kondo lattice model, we conclude that the Fermi surface will correspond to an electron density of $1+(1-p) = 2-p$: this is a small Fermi surface of holes of density $p$. The second layer of $\widetilde{f}$ spins forms a spin liquid with fractionalized spinon excitations. The consistency of this structure with extended Luttinger relations on Fermi surface size has been discussed in Ref.~\cite{nikolaenko2021}.

As in the theory of the heavy Fermi liquid, the FL* phase here is a phase in which all gauge symmetries associated with the Kondo coupling $J_K$ are fully broken by a Higgs condensate. Here, we identify the condensate as the mean field order parameter \cite{Zhang2020}
\beq
\phi_i \propto - J_K \langle f^\dagger_{i\alpha} c_{i \alpha} \rangle
\eeq
With this condensate, we can write down a simple effective Hamiltonian of the electronic excitations of the FL* state \cite{Zhang2020}
\begin{align}
H=-  \sum_{i,j} t_{ij} c^\dagger_{i;\alpha} c_{j;\alpha}+  \sum_{i,j} t_{1,ij} f^\dagger_{i;\alpha} f_{j;\alpha}+ \sum_{i}\phi_i(c^\dagger_{i;\alpha} f_{i;\alpha}+f^\dagger_{i;\alpha} c_{i;\alpha})
\label{eq:initial_Hamiltonian}
\end{align}
In our analysis below, $t_{1,ij}$ and $\phi_i$ will be treated as adjustable parameters. 

The theory for the FL* state also contains the $\widetilde{f}$ layer.
The gauge symmetries associated with $\widetilde{f}$ layer remains at least partially unbroken, and are needed to realize the fractionalized spinon excitations of the FL* state. Some earlier discussions of the FL* phase \cite{YQSS10,EGMSS11} or the pseudogap metal \cite{YRZ,YRZ_rev} have similarities to a model obtained by adding a single `hidden' band near half-filling.
The important spectra of fractionalized spin excitations associated with the second hidden layer are realized in Refs.~\cite{YQSS10,EGMSS11} by bosonic spinons, while they appear to be absent in Refs.~\cite{YRZ,YRZ_rev}. The presence of these fractionalized spin excitations 
in the second layer of hidden spins produced by fractionalizing the paramagnon is one of the key distinguishing features of our model of the pseudogap metal, and we discuss their consequences for the electron spectrum in Section~\ref{sec:frac}.


\subsection{Electron self energy from
couplings to the second hidden layer}
\label{sec:frac}

This subsection turns to a computation of the electronic self-energy in the paramagnon fractionalization theory \cite{Zhang2020,Zhang2020_1,nikolaenko2021}. In the pseudogap metal phase, the gauge symmetry associated with the first hidden layer is higgsed, and so the only gapless modes in the $c$ and $f$ layers are those associated with the electronic Fermi surface (see Fig.~\ref{fig:layers}). The self energy from these modes will therefore be Fermi liquid-like. However, there can be gapless excitations on the second hidden layer of ${\bm S}_2$ spins, which will exchange couple to the electronic Fermi surface, and yield a damping beyond that is present in a Fermi liquid. This layer of $\widetilde{f}$ fermions   has so far played no direct role in our computations. To second order in $J_\perp$, the second layer contributes a $f$ fermion self energy
\beq
\Sigma_{f} (k,i\omega_n) = \frac{3J_\perp^2}{4\beta} \int \frac{d^2q}{(2\pi)^2}\sum_{\nu_m} \chi(q,i\nu_m) G_{f}(k-q,i\omega_n-i\nu_m) \,.
\label{sigmaf}
\eeq
Similarly, the coupling in Eq.~(\ref{tJKval}) leads to a self energy for the $c$ fermions
\beq
\Sigma_{c} (k,i\omega_n) = \frac{3\widetilde{J}_K^2}{4\beta} \int \frac{d^2q}{(2\pi)^2}\sum_{\nu_m} \chi(q,i\nu_m) G_{c}(k-q,i\omega_n-i\nu_m) \,.
\label{sigmac}
\eeq
Here $\chi$ is the spin susceptibility of second hidden layer
\be
\chi (q,i\nu_m) = \frac{1}{3} \int_0^\beta d\tau \left\langle T \left[ {\bf S}_2 (q,\tau) \cdot {\bf S}_2 (-q,0) \right] \right\rangle e^{i\nu_m \tau}.
\ee
These self-energies are dependent upon the unknown spin susceptibility of the spin liquid on the second hidden layer. In Appendix~\ref{app:z2} we will consider a realistic candidate gapless $\mathbb{Z}_2$ spin liquid which has been the focus of significant recent interest \cite{TSMPAF99,WenPSG,Becca01,SenthilIvanov,SenthilLee05,Kitaev2006,Becca13,Sandvik18,Becca18,Becca20,Imada20,Gu20,shackleton2021,Gu2021}. The needed computations are numerically demanding, and we compute the $f$ and $c$ self energies for representative parameters. 

In the present section we will examine a simple approximation for $\chi$, in which we replace it by the momentum-independent spin susceptibility of the SYK model \cite{SY,Parcollet1,CGPS}
\beq\label{eq:SY_suscept}
\chi (q, \tau) &\sim & \frac{T}{\sin (\pi T \tau)}; \nonumber \\
\chi (i \nu_n) &=& \int_{-\infty}^{\infty} \frac{ d \omega}{\pi} \frac{\rho_\chi (\omega)}{\omega - i \nu_n}; \nonumber \\
\rho_\chi(\omega) &=& -\frac{X}\pi \tanh\left(\frac{\omega}{2T}\right)\,,
\eeq
for some constant $X$. Inserting this susceptibility in Eqs.~(\ref{sigmaf}) and (\ref{sigmac}) we see that the fermion self energies also become momentum independent. Furthermore, if we assume that the local fermion density of states
\be
\rho_f (\omega) = - \frac{1}{\pi} \int \frac{d^2 k}{(2 \pi)^2} \mbox{Im}\, G_f (\omega, k)
\ee
(and similarly for $\rho_c (\omega)$) is $\omega$ independent at small $\omega$, then Eqs.~(\ref{sigmaf}) and (\ref{sigmac}) evaluate to the marginal Fermi liquid self energy in Eq.~(\ref{eqn:marginal_self_energy}). Below, we present computations in which we do {\it not\/} assume $\omega$-independence in $\rho_{f,c} (\omega)$, and evaluate the resulting self energies in Eqs.~(\ref{sigmaf}) and (\ref{sigmac}).

Inserting the momentum-independent spectral density in $\rho_\chi$ into Eq.~(\ref{sigmaf}), and analytically continuing to the real frequency axis, and taking the imaginary part we obtain
\beq
\Sigma''_{f}(\omega) = \frac{3 J_\perp^2}{4\pi} \int_{-\infty}^{+\infty} d\omega_1 \rho_\chi(\omega_1)\rho_{f}(\omega - \omega_1) (n_F(\omega_1 - \omega)n_B(-\omega_1 )-n_F(\omega - \omega_1)n_B(\omega_1 ))\,.
\eeq
This can be simplified and written as
\beq
\Sigma''_{f}(\omega) = \frac{3 J_\perp^2}{4\pi} \int_{-\infty}^{+\infty} d\epsilon \rho_\chi(\omega - \epsilon)\rho_{f}(\epsilon) (n_F(\epsilon) +n_B(\epsilon- \omega)). \label{eq:Sigmaf}
\eeq
The real part can be obtained using the Kramers-Kronig relation from the imaginary part
\beq\label{eq:SelfEnf}
\Sigma'_{f}(\omega) = \int_{-\infty}^{+\infty} \frac{d\nu}\pi \frac{\Sigma''(\nu) - \Sigma''(\omega)}{\nu-\omega}.
\eeq

On the right-hand-side of Eq.~(\ref{eq:Sigmaf}) we will insert the spectral density $\rho_f$ of the $f$ fermions which does not include the contribution of $\Sigma_f$. This is obtained from Green's function of the $f$ fermions which does include the hybridization with conduction electrons 
\beq\label{eq:Gpsi1}
G_{f}(i\omega_n,k) = \frac1{i\omega_n - \epsilon_f(k) - \phi^2 G_c^0(i\omega_n,k) }
\eeq
where $\phi$ is a hybridization constant and
\beq\label{eq:Gc0}
G_c^0(i\omega_n,k) = \frac1{i\omega_n - \epsilon_c(k)}
\eeq
is the bare Green's function for the $c$ electron layer. Taking the imaginary part of the analytically continued Green's function \eqref{eq:Gpsi1} gives the spectral function 
\beq\label{eq:rhof}
\rho_{f}(\omega) =  \int \frac{d^2k}{(2\pi)^2} \left(v_k^2 \delta(\omega - \xi^+_k) + u_k^2 \delta(\omega - \xi^-_k)\right)
\eeq
where we introduced some momentum-dependent functions
\beq\label{ukvk}
u_k^2 = \frac12\left(1+\frac{\psi_k}{\sqrt{\psi_k^2 + \phi^2}}\right),\,\,\,\,v_k^2 = \frac12\left(1-\frac{\psi_k}{\sqrt{\psi_k^2 + \phi^2}}\right),
\eeq
\beq\label{xik}
\psi_k = \frac{\epsilon_c(k) - \epsilon_f(k)}{2},\,\,\,\xi_k^\pm= \frac{\epsilon_c(k) + \epsilon_f(k)}{2} \pm \sqrt{\psi_k^2 + \phi^2}.
\eeq

The self-energy for the $c$ electrons has a similar form
\beq\label{eq:SelfEnc}
\Sigma''_{c}(\omega) = \frac{3 \tilde{J}_K^2}{4\pi} \int_{-\infty}^{+\infty} d\epsilon \rho_\chi(\omega - \epsilon)\rho_{c}(\epsilon) (n_F(\epsilon) +n_B(\epsilon- \omega)). \label{eq:Sigmac}
\eeq
with the $c$ electron spectral function $\rho_c(\epsilon)$ obtained from Eq.~(\ref{eq:Gc0}). 

Results from computation of the above expressions appear in Section~\ref{sec:2212}, where we will compare them to the momentum and energy distribution curves (MDC and EDC) from ARPES observations on Bi2212.

\section{B\lowercase{i}2201}
\label{sec:2201}

We first focus on modeling photoemission observations \cite{He2011} in the {\it monolayer\/} cuprate Bi2201 (Pb$_{0.55}$Bi$_{1.5}$Sr$_{1.6}$La$_{0.4}$CuO$_{6+\delta}$) and compare it to the paramagnon fractionalization theory for a single layer Hubbard model. 

On the overdoped side, $\phi=0$ and the physical layer decouples from the hidden layers. Then, the system is described by a tight-binding Hamiltonian and hopping parameters can be obtained by fitting a band structure. The band structure is derived from the maxima of ARPES EDC (energy distribution curves), and the following set of parameters fits the experiment \cite{He2011}: 
\bea
\epsilon_c(k) &=&-2t(\cos k_x+\cos k_y)-4 t' \cos k_x \cos k_y-2t''(\cos 2 k_x+\cos 2 k_y) \nonumber \\
&~&~~~-4t'''(\cos 2 k_x \cos k_y+\cos 2 k_y \cos k_x)-\mu_c\,,
\eea
where $t=0.22, t'=-0.034, t''=0.036, t'''=-0.007, \mu_c=-0.24$ (all energies are in units of eV). 

On the underdoped side, we have $\phi \neq 0$, and the physical layer hybridizes with the first hidden layer. The dispersion of hybridized bands from the Hamiltonian Eq.~(\ref{eq:initial_Hamiltonian}) is:
\be
E_{\pm}(k)=\frac{\epsilon_c(k)+\epsilon_f(k)}{2} \pm \sqrt{\left(\frac{\epsilon_c(k)-\epsilon_f(k)}{2}\right)^2+\phi^2 }
\label{eq:Epm}
\ee
The salient feature of cuprate superconductors is the emergence of hole pockets in the ARPES spectra. We choose tight-binding parameters for the hidden layer that captures this feature and fix the number of particles in both layers: $c_{i \alpha} ^\dag c_{i \alpha} =(1-p)/2$, where $p$ is the doping, and $f_{i\alpha}^\dag f_{i\alpha}=1/2$. We assumed the hidden layer dispersion
\beq
\epsilon_f(k) = 2t_1(\cos k_x+\cos k_y)+4 t_1' \cos k_x \cos k_y +2t_1''(\cos 2 k_x+\cos 2 k_y)-\mu_f\,, \label{eq:ef}
\eeq
where $t_1=0.1, t_1'=-0.03, t_1''=-0.01, \mu_f=0.009$. Finally, we use the hybridization $\phi=0.09$ and temperature $T=40$K, which sets the size of the gap near the antinodes. 

\begin{figure}
\begin{minipage}[h][0.4\linewidth][t]{0.38\linewidth}
  \center{\includegraphics[width=1\linewidth,height=1\textwidth]{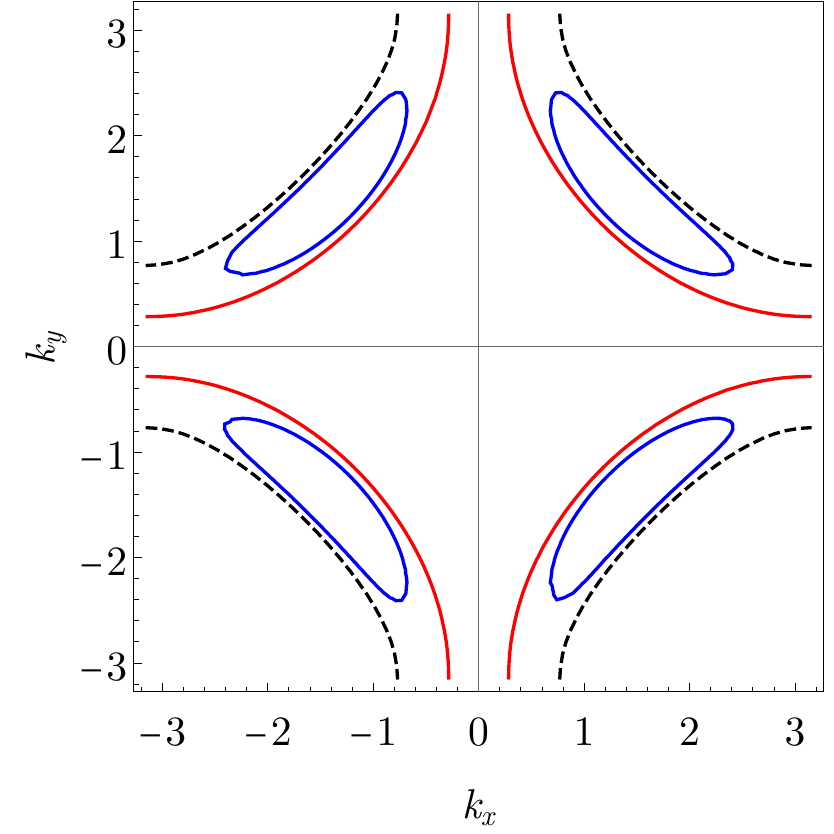}}
  a)
  \end{minipage} 
  ~~~~~
    \begin{minipage}[h][0.4\linewidth][t]{0.55\linewidth}
    \center{\includegraphics[height=0.7\linewidth,width=1\linewidth]{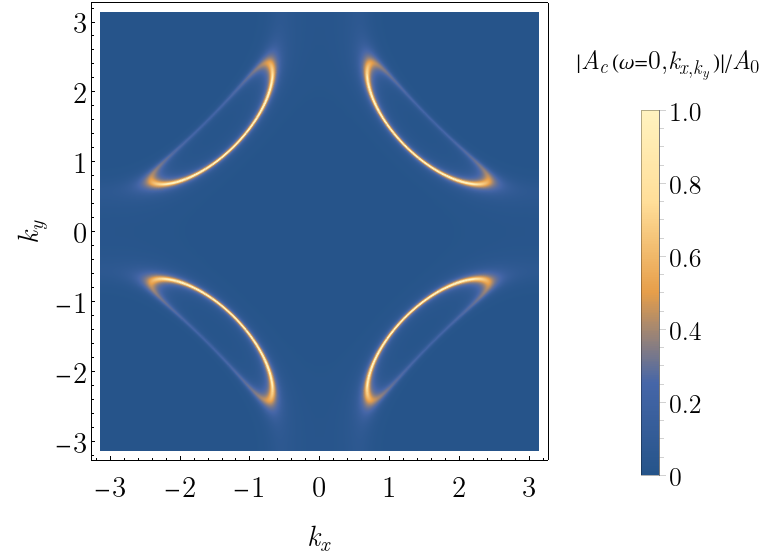}}
     \\b) 
     \end{minipage}
\caption{a) Fermi surface of unhybridized and hybridized bands. Red line shows the physical layer band ($\epsilon_c(k)=0$), black dashed line shows the first hidden layer band ($\epsilon_f(k)=0$), while the blue lines is the final hybridized Fermi surface ($E_{\pm}(k)=0$). (b) Spectral function $A_{cc}({\bf k}, \omega=0)$ of the $c$-electrons at the Fermi energy obtained with $\Sigma_c=\Sigma_f=-i 0.005 eV$.}
\label{fig:brillouin_zone}
\end{figure}

In Figure~\ref{fig:brillouin_zone}(a) we present the resulting Fermi surface in the first Brillouin zone, exhibiting hole pockets, while in Figure~\ref{fig:brillouin_zone}(b) we plot the corresponding spectral function of the $c$-electrons, $A_{cc}({\bf k}, \omega=0)$, which reflects the ARPES intensity and is defined via 
\begin{equation}
    A_{cc}({\bf k},\omega)=-\frac{1}{\pi} \text{Im}\left( \frac{\omega-\epsilon_f-\Sigma_f}{(\omega-\epsilon_c-\Sigma_c)(\omega-\epsilon_f-\Sigma_f)-\phi^2}\right)\,.
\label{eqn:spectral_functions}
\end{equation}
where  $\Sigma_c,\Sigma_f$ represent a disorder induced broadening. In agreement with ARPES experiments, we find that the ARPES intensity is largely confined to the sides of the hole pockets that are closest to the $\Gamma$ point, and vanishes on the opposite sides.
We note that the quasiparticle residue, which modulates the ARPES intensity, does not contribute to transport observables, and so these hole pockets can serve as an explanation for recent angle-dependent magnetoresistance measurements \cite{ADMR}.

\begin{figure}
\begin{minipage}[h]{0.75\linewidth}
  \center{\includegraphics[width=1\linewidth]{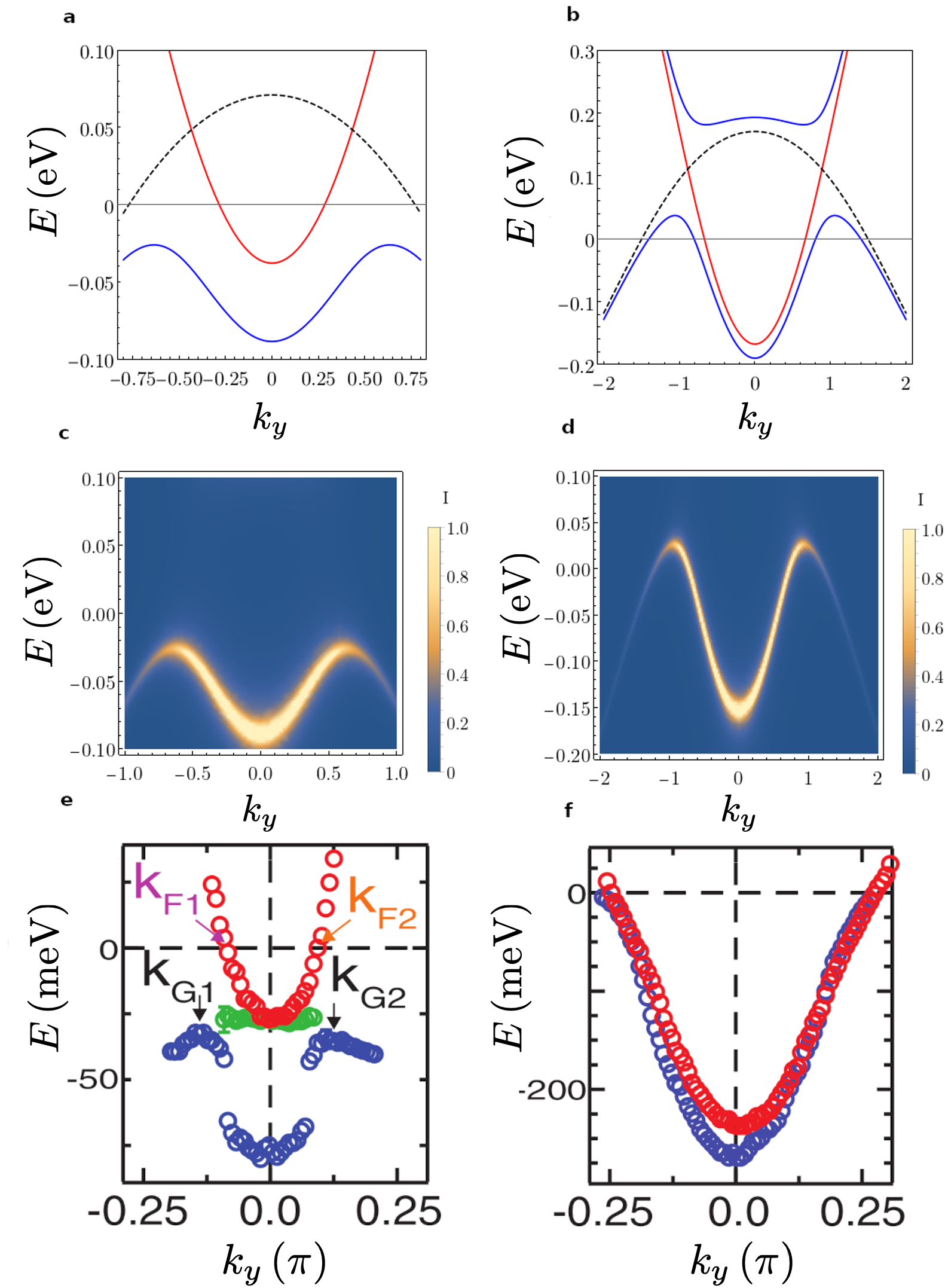}}
  \end{minipage} 
\caption{Antinodal (\textbf{a}) and nodal (\textbf{b}) bands along constant $k_x$. Red line shows physical layer $c$ band, black dashed line shows the first hidden layer $f$ band while blue lines are hybridized dispersions. (\textbf{c}), (\textbf{d}) Theoretically computed ARPES intensity ($c$-electron spectral function) at constant $k_x$ in arbitrary units in the anti-nodal and nodal regions. While the antinodal spectrum shows the pseudogap, the nodal spectrum coincides with the overdoped case. (\textbf{e}), (\textbf{f}) ARPES observations from He {\it et al.} \cite{He2011} of the band structure in the underdoped (blue dots) and overdoped (red dots) cases, in the antinodal and nodal regions. The data is obtained from the maxima of the EDC curves at different momenta.}
\label{fig:He}
\end{figure}
Next, we consider the spectrum of gapped excitations near the antinode, and to this end plot in Fig.~\ref{fig:He}(a)
the band dispersion along $k_x=-\pi$. We find that the hybridized band (blue line) has a Bogoliubov-like shape, similar to ordinary superconductors. An important difference, however, is that the minimum in the excitation energy of the blue band occurs at a momentum that is slightly shifted from the unhybridized Fermi momenta. This occurs because the system does not have particle-hole symmetry, which distinguishes the paramagnon fractionalization model from an ordinary superconducting model.  

Once we move away from the antinodal point, the bottom band rises, the gap closes, and hole pockets appear, as shown in Fig.~\ref{fig:He}(b) for $k_x=-2$. Contrary to the antinodal region, the hybridized band nearly coincides with the unhybridized band near the Fermi surface. As mentioned above, the sides of the hole pockets facing $(\pi,\pi)$ cannot be detected by ARPES experiments as most of the spectral weight remains concentrated in the parts of the original unhybridized band facing the $\Gamma$-point (see Fig.~\ref{fig:brillouin_zone}(b)). 
In Figs.~\ref{fig:He}(c),(d) we present the corresponding spectral function, $A_{cc}({\bf k}, \omega)$, again showing that the ARPES intensity on the parts of the hole pockets facing the $(\pi,\pi)$-point is largely diminished, in agreement with Fig.~\ref{fig:brillouin_zone}(b).
For comparison, we present the experimentally measured ARPES intensity by He {\it et al.} \cite{He2011} in Figs.~\ref{fig:He}(e),(f), which for the underdoped compounds, agrees well with our theoretical results shown in Figs.~\ref{fig:He}(c),(d). The experimental data also reveal 
a sharp distinction between the antinodal and nodal regions:  while in the antinodal region, the electronic band for the underdoped compound lies below that of the overdoped cuprate, and possesses a Bogoliubov-like shape, in the nodal region the bands in the underdoped and overdoped compounds coincide.

Appendix~\ref{app:sdwpdw} contains a brief discussion of other models of the underdoped pseudogap metal: we compare corresponding results in the SDW and PDW models with those of the paramagnon fractionalization theory.

\subsection{Comparison with STM experiments}
\label{sec:stm}

STM experiments on several BSCCO compounds \cite{Fischer2007}, and in particular on Bi2201 \cite{Kugler2001}, have reported a gradual evolution of the superconducting state into the pseudogap state upon raising the temperature. The familiar V-shaped STM profile from $d$-wave superconductivity gradually evolves into the Fano-like profile. We show below that the paramagnon fractionalization model yields a Fano-like lineshape similar to STM observations. 

\begin{figure}
\begin{minipage}[h]{0.4\linewidth}
    \raggedright ~~a) 
  \center{\includegraphics[width=1\linewidth]{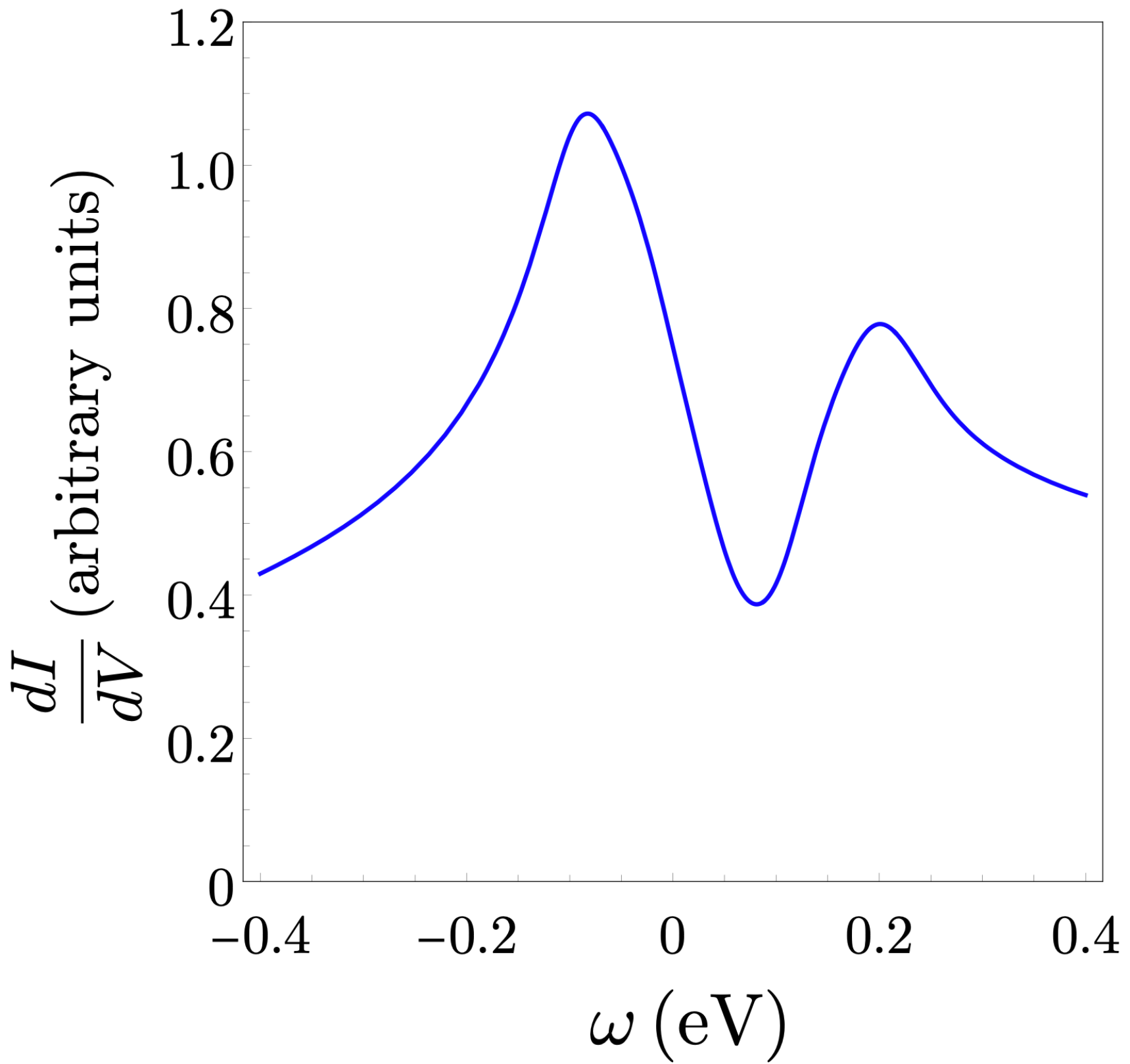}}
  \end{minipage} 
    ~~~~~
    \begin{minipage}[h]{0.4\linewidth}
     \raggedright ~~b)
    \center{\includegraphics[width=1\linewidth]{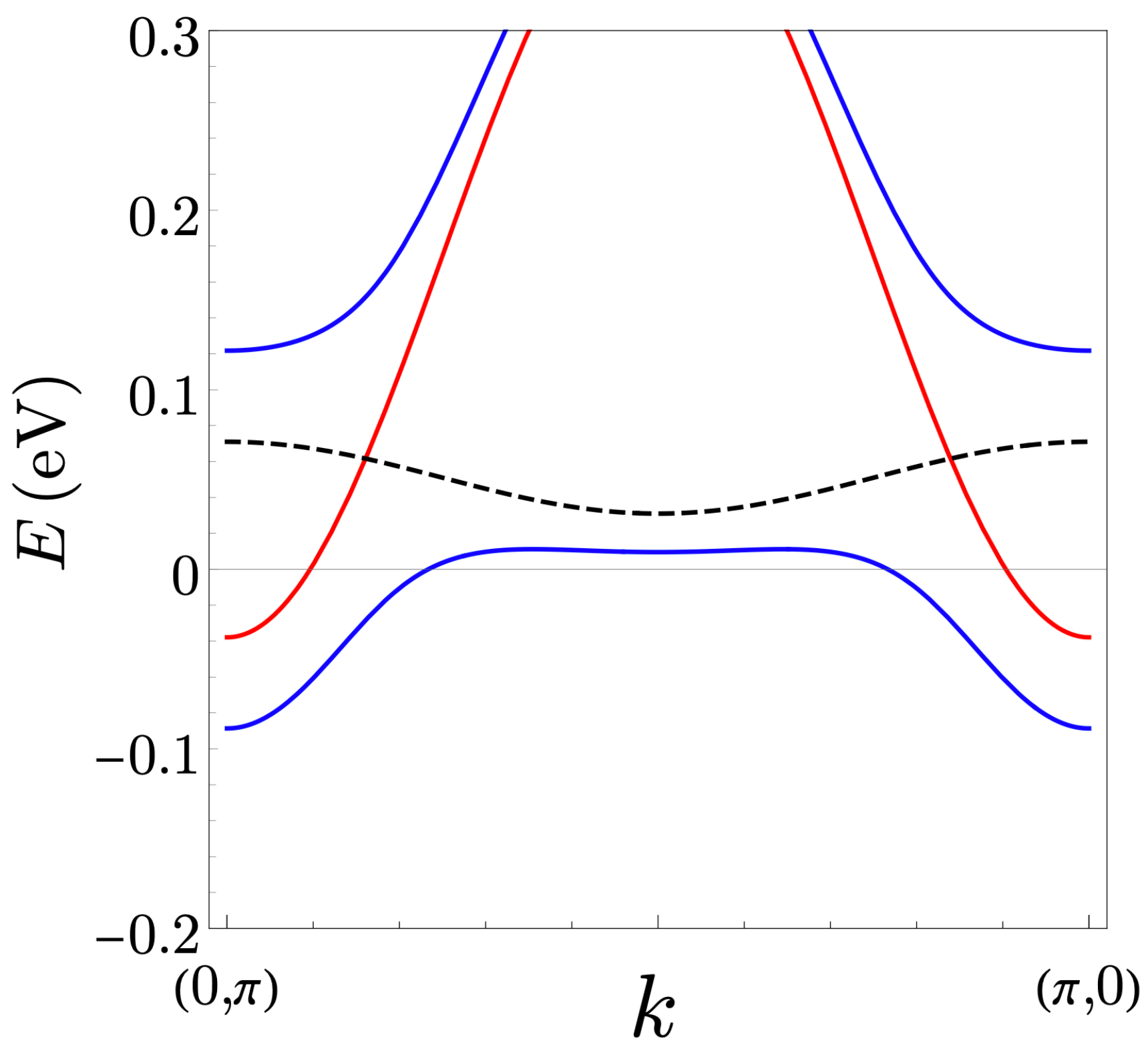}}
    \end{minipage} 
  \caption{a) STM spectra of the paramagnon fractionalization model with broadening $\Gamma_0=0.04$. b) 1-d bands cut from $(k_x,k_y)=(0, \pi)$ to $(k_x,k_y)=(\pi, 0)$. The tight-binding parameters are same as in Fig.~\ref{fig:brillouin_zone}.}
  \label{fig:stm}
\end{figure}

The differential conductance, $dI/dV$, can be obtained via \cite{Dirk2017}
\begin{equation}
    \frac{dI}{dV}(r,\omega)=-\frac{e}{\hbar}N_t\sum_\alpha\sum_{i,j=1}^2 \left[\hat{t} \,\im G^R(r,r,\omega) \hat{t}\right]_{ij} \ ,
    \label{eqn:differential_conductance}
\end{equation}
where $\hat{t}$ is the diagonal tunneling matrix, representing the tunneling of electrons between the STM tip and the $c$- and $f$-bands in the system. In a translationally invariant system with no defects, one obtains
\begin{equation}
    \frac{dI}{dV}(r,\omega)=\frac{e \pi}{\hbar}N_t\sum_{k,\alpha} \left(t_c^2 A_{cc}(\omega,k)+2t_c t_f A_{cf}(\omega,k)+t_f^2 A_{ff}(\omega,k)\right) \ .
\end{equation}
For simplicity, we assume $t_f=0$, implying that electrons can tunnel only into the physical layer. The resulting $dI/dV$ is shown in Fig.~\ref{fig:stm}(a).
The minimum of the $dI/dV$ curve occurs at a positive energy, and arises from the gap between the upper and lower electronic bands. The dispersions of the bands are given by $E_\pm({\bf k})$ (see Eq~(\ref{eq:Epm})), as shown in 
Fig.~\ref{fig:stm}(b). Since the broadening $\Gamma_0$ is of the order of the gap, the differential conductance has a Fano-like shape instead of a hard gap. A similar $dI/dV$ lineshape with the shifted minima was observed in the STM measurements \cite{Davis09} at low temperatures. 

\section{B\lowercase{i}2212}
\label{sec:2212}

\begin{figure}
\includegraphics[width=.48\linewidth]{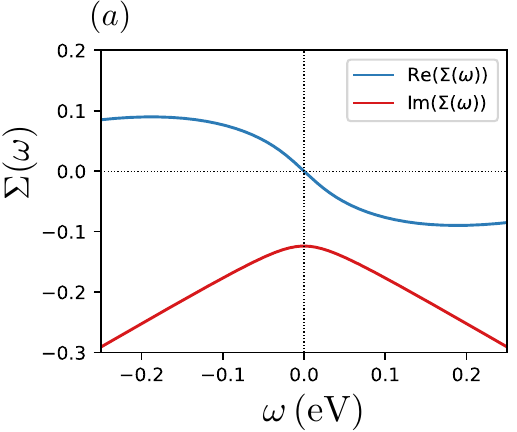}\,\,
\includegraphics[width=.48\linewidth]{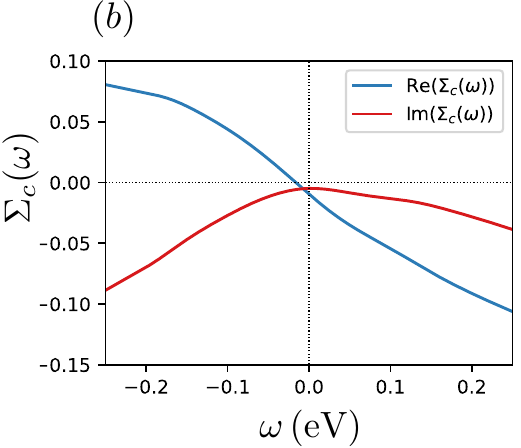}\,\,
\includegraphics[width=.48\linewidth]{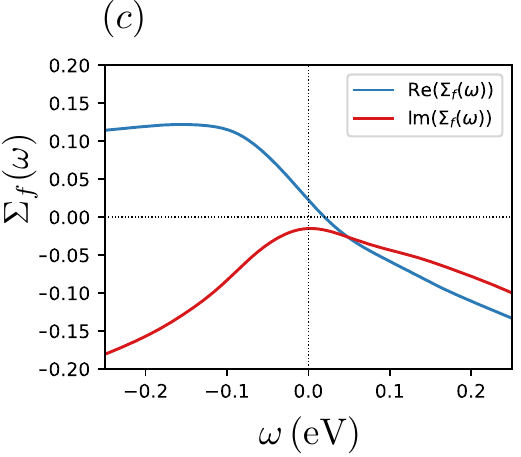}
\caption{Real (blue lines) and imaginary (red lines) parts of self energies for (a) marginal Fermi liquid \eqref{eqn:marginal_self_energy} with parameters $\lambda=\omega_c = 0.5 $ and $\Gamma_0 = 0$, (b) $c$ electrons with self energy from the SYK spin liquid with $\tilde{J}_K^2 = 4\pi/3$ from Eq.~(\ref{eq:Sigmac}) and (c) $f$ fermions  with self energy from the SYK spin liquid with $ J_\perp^2 = 4\pi/3$ from Eq.~(\ref{eq:Sigmaf}) . Temperature is $T=150$K.}
\label{fig:SelfEn}
\end{figure}

We next turn to (Bi,Pb)$_2$Sr$_2$CaCu$_2$O$_{8+\delta}$ (Bi2212), which is a {\it bilayer} compound possessing two CuO planes per unit cell. We therefore need to consider two corresponding hidden layers attached to each CuO layer. We assume the following tight binding dispersion  \cite{drozdov2018phase,Chakravarty1993}: 
\beq\label{eq:epsc}
\epsilon_c(k)=-2t(\cos k_x+\cos k_y)-4 t' \cos k_x \cos k_y-2t''(\cos 2 k_x+\cos 2 k_y)-\mu_c \pm \epsilon_{\perp} (k) \,, 
\eeq
with $t=0.36, t'=-0.108, t''=0.036, t_{\perp}=0.108$, $\mu_c=-0.4$ and  $\epsilon_{\perp} (k)  =t_{\perp}/4 (\cos k_x-\cos k_y)^2$ describes the coupling between the two layers,  

In order to describe the EDC and MDC spectra measured in ARPES experiments by Chen {\it et al.} \cite{chen2019incoherent} on Bi2212, it is necessary to consider the electronic self-energy arising within our model, which we compute in Section~\ref{sec:frac} and Appendix~\ref{app:z2}. 
A simple model of this self energy is that of a marginal Fermi liquid, given by 
\begin{equation}
    \Sigma_c(\omega)=\lambda\left(\omega \ln \frac{x}{\omega_c}-i \frac{\pi}{2}x\right)-i \Gamma_0 , \,\,\,\,x=\sqrt{\omega^2+\pi^2 T^2} 
    \label{eqn:marginal_self_energy}
\end{equation}
which was originally proposed on phenomenological grounds \cite{Varma1989}. We show its behavior in Fig.~\ref{fig:SelfEn}(a). As we noted in Section~\ref{sec:frac} such a marginal Fermi liquid self energy emerges from an SYK model for the spinon fluctuations in the second hidden layer, when we neglect the energy dependence of the local electron density of states. A model which does not ignore the energy dependence of the electron density of states was presented in Section~\ref{sec:frac} and behavior of the self energies for $f$ and $c$ fermions is shown in Fig.~\ref{fig:SelfEn}(b),(c), and we will also apply its results below. As we will show, this model improves the agreement with experiments. 

\begin{figure}
\begin{minipage}[h]{0.45\linewidth}
  \center{\includegraphics[width=1\linewidth]{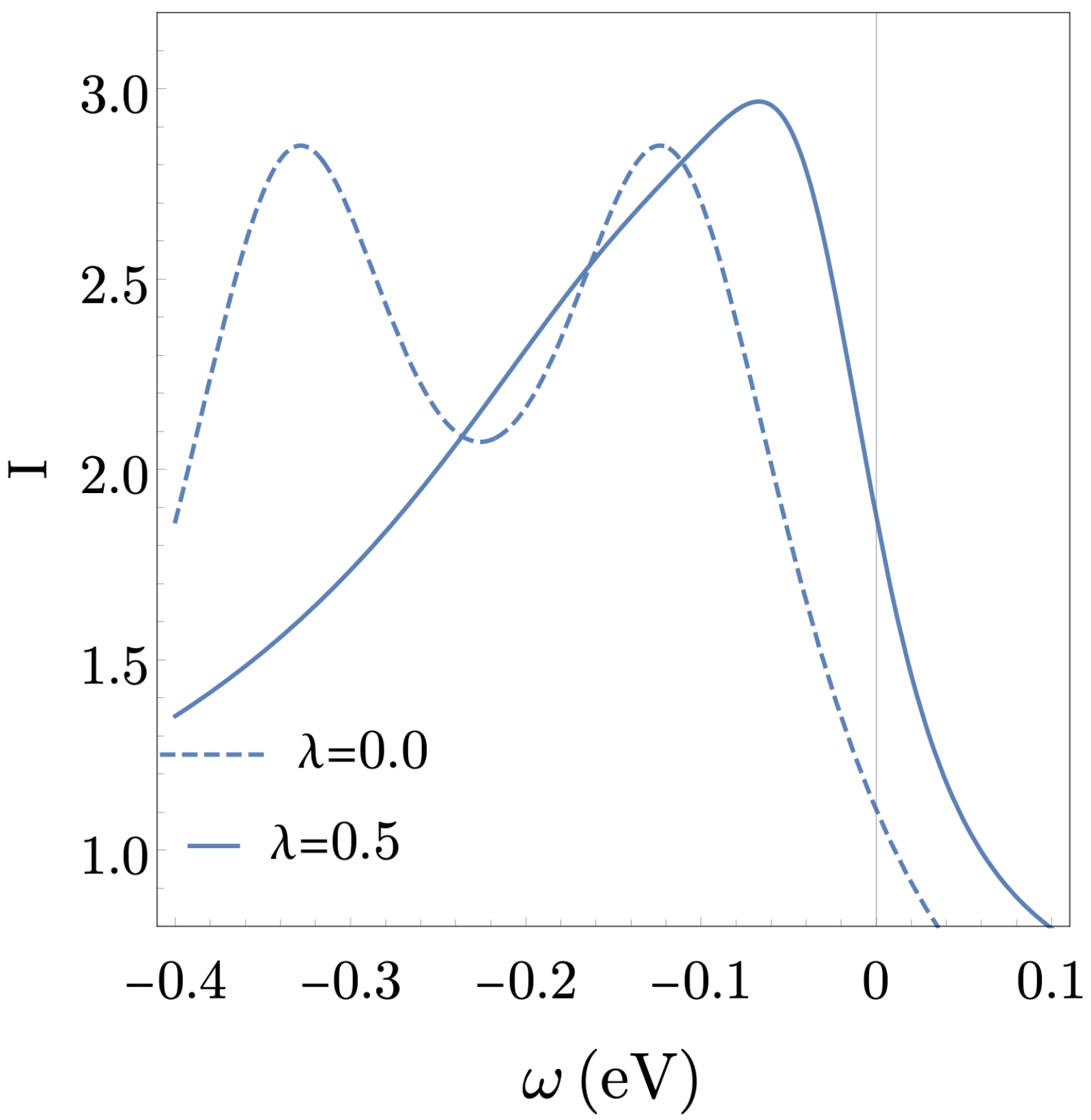}}
 \\a)EDC cut along $k_y=0$.
  \end{minipage} 
  ~~~~~  
    \begin{minipage}[h]{0.45\linewidth}
    \center{\includegraphics[width=1\linewidth]{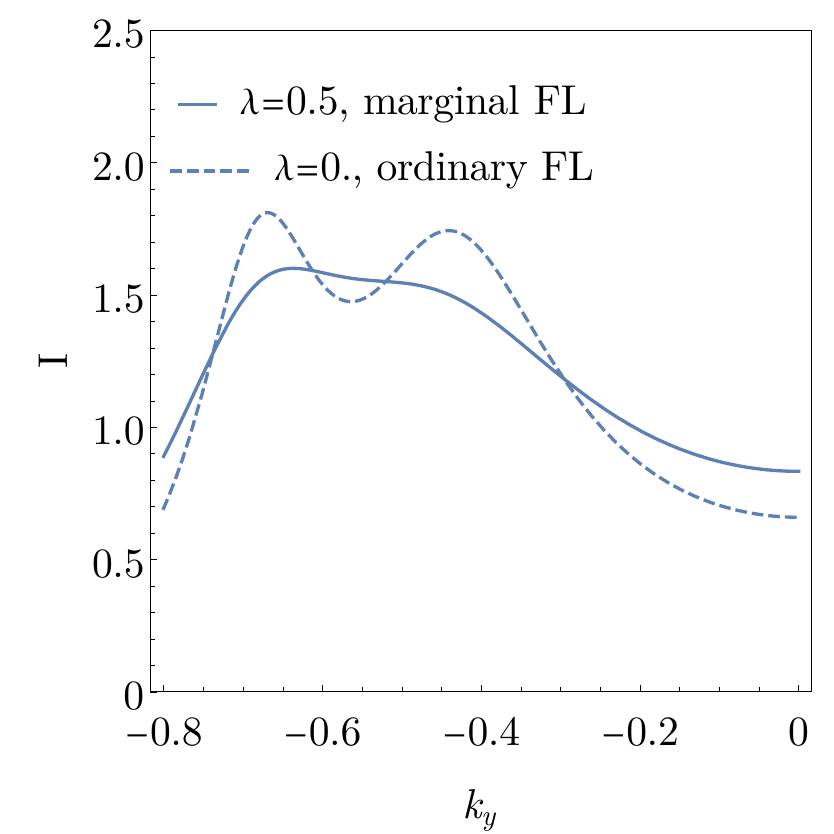}}
     \\b)MDC cut at $\omega=0$.
    \end{minipage} 
\caption{ARPES computations near the antinodal point without hidden layers for $T=250$K.
These are to be compared with experimental observations in the OD86 overdoped regime shown in Fig.~\ref{fig:chen_experiment}(e,f).
Solid lines correspond to marginal Fermi liquid with $\lambda=0.5$, while dashed curves correspond to ordinary Fermi liquid with $\lambda=0$. Here and below we normalize EDC and MDC curves to have same area at the chosen region. Here, $\omega_c=0.5$ and $\Gamma_0=0.092$ from \cite{chen2019incoherent}.}
 \label{fig:chen_fl_cut}
\end{figure}

\begin{figure}
\begin{minipage}[h]{1\linewidth}
  \center{\includegraphics[width=.8\linewidth]{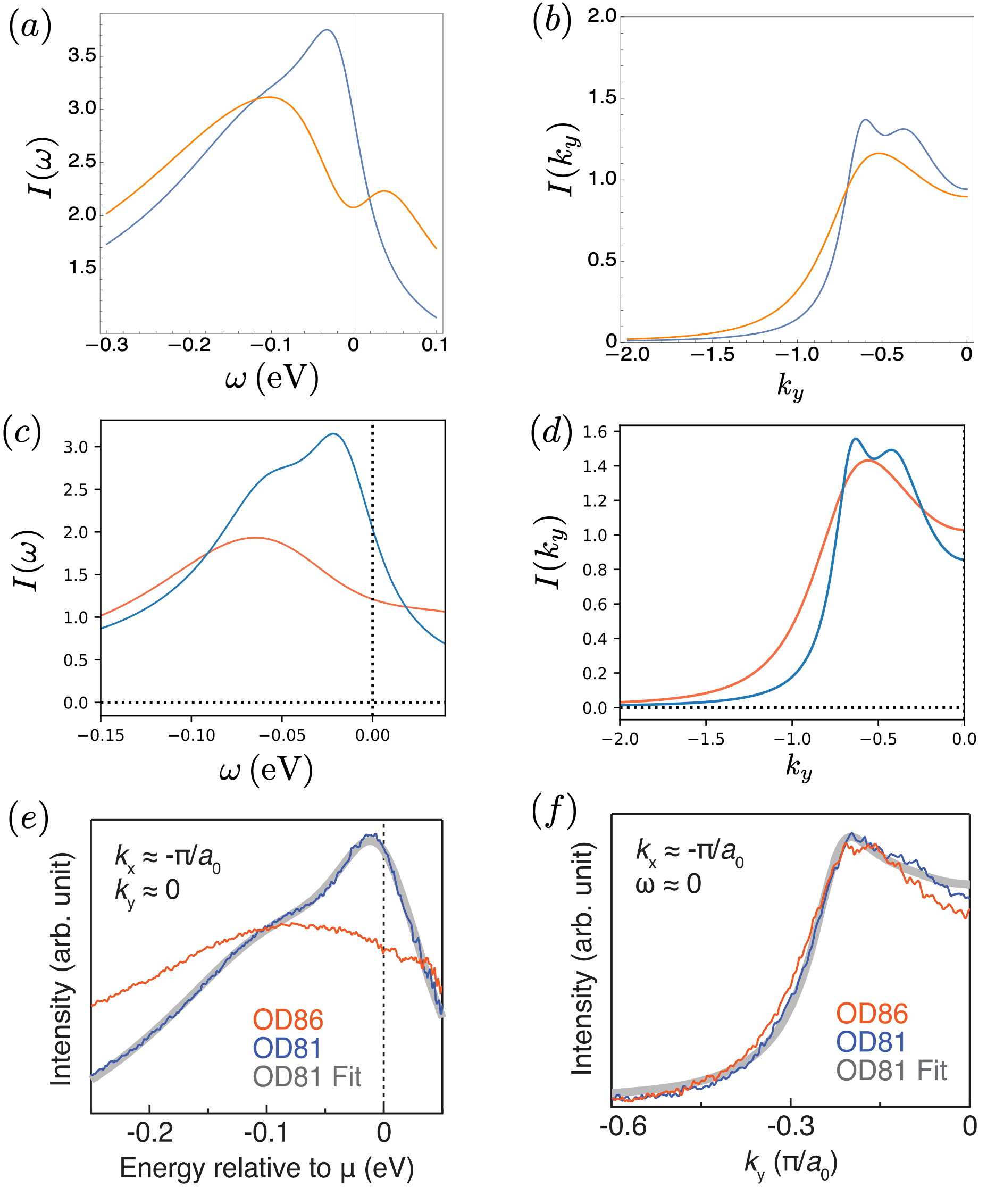}}
  \end{minipage}
\caption{(\textbf{a}), (\textbf{b}) Computations of ARPES EDC and MDC curves in the antinodal region for $T=150$K and other parameters given above. Blue line correspond to the overdoped case, while orange line correspond to an underdoped case when the physical layers are coupled to the hidden layers with $\phi=0.15$.  We use the MFL self energy in Eq.~(\ref{eqn:marginal_self_energy}) with $\lambda = 0.5$, $\omega_c=0.5$ and $\Gamma_0=0.092$. (\textbf{c}), (\textbf{d}): same curves as above but with self energies from the SYK spin liquid in \eqref{eq:SelfEnf} and \eqref{eq:SelfEnc}. The coupling constants are $J_\perp = 1$, $\tilde{J}_K = 5$ and the SY parameter is chosen $X = 0.5$. The value of the impurities-induced broadening is $\Gamma_0 = 0.09$. (\textbf{e}), (\textbf{f}) Experimental ARPES EDC and MDC curves from Ref.~\cite{chen2019incoherent}. }
\label{fig:chen_experiment}
\end{figure}

We begin by considering the case where we neglect the hidden layers, as would be appropriate in the overdoped regime, and present the resulting
EDCs and MDCs near the antinodal point in Fig.~\ref{fig:chen_fl_cut};
these are to be compared with the experimental observations shown later in Fig.~\ref{fig:chen_experiment}.
Due to the bilayer splitting, the corresponding EDC and MDC curves with $\lambda=0$ (dashed lines) exhibit two peaks.  These two peaks exhibit the same width, due to the energy independent disorder-induced broadening $\Gamma_0$. However, experimentally, only a single peak is observed (see Fig.~\ref{fig:chen_experiment}a), which we also obtain within our model when including the strongly frequency dependent marginal Fermi liquid self-energy of Eq.~(\ref{eqn:marginal_self_energy}).

Next, we turn to the underdoped regime, where there are two additional hidden layers, which we choose to have the same dispersion as in Eq.~(\ref{eq:ef}) with $t_1=0.108, t_1'=0.0, t_1''=-0.01, \mu_f=-0.06$ and hybridization $\phi=0.15$. The Hamiltonian matrix for this system in the basis $\ket{1_a,1,2,2_a}$, where $a$ denotes the hidden layer, is given by 
\begin{equation}
    {\hat H}=\left(
    \begin{array}{cccc}
    \epsilon_f     &\phi & 0 & 0   \\
    \phi     &  \epsilon_c & \epsilon_\perp & 0\\
    0 & \epsilon_\perp &\epsilon_c & \phi\\
    0 & 0 & \phi & \epsilon_f
    \end{array}
    \right)
\end{equation}
The resulting spectral function for the physical $c$-electrons is given by 
\begin{equation}
    A_{cc}=-\frac{1}{\pi} \text{Im}\left( \frac{\omega-\epsilon_f-\Sigma_f}{(\omega-\epsilon_c-\epsilon_{\perp}-\Sigma_c)(\omega-\epsilon_f-\Sigma_f)-\phi^2}+\frac{\omega-\epsilon_f-\Sigma_f}{(\omega-\epsilon_c+\epsilon_{\perp}-\Sigma_c)(\omega-\epsilon_f-\Sigma_f)-\phi^2}\right)
\end{equation}
We take a MFL form in Eq.~(\ref{eqn:marginal_self_energy}) for the self-energies in the physical and hidden layers.
\begin{figure}
\begin{minipage}[h]{0.4\linewidth}
  \center{\includegraphics[width=1\linewidth]{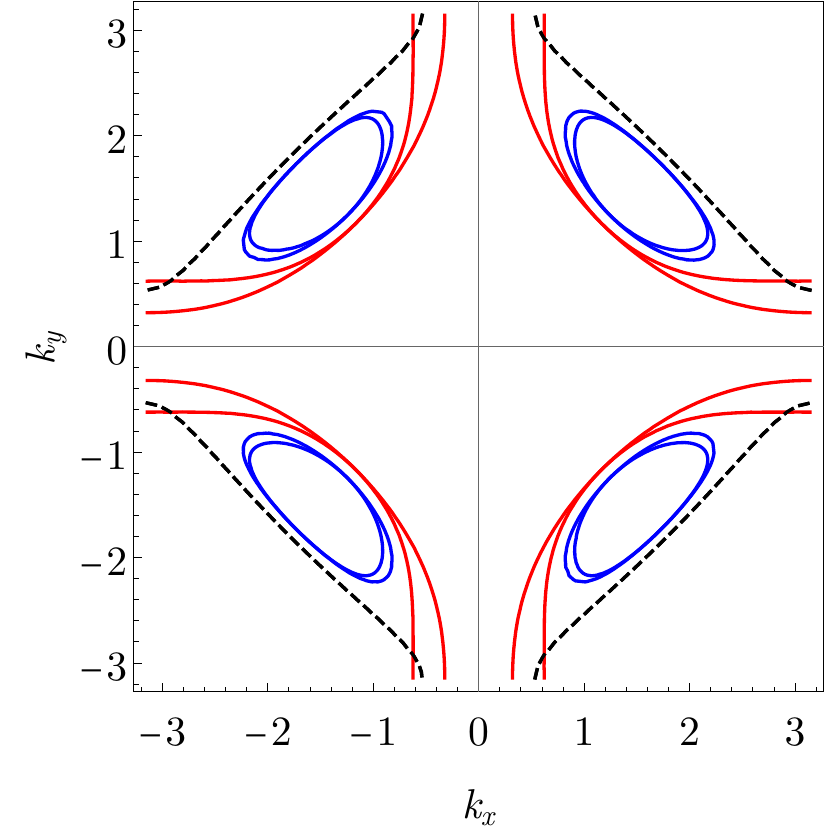}}
 \\a)
  \end{minipage} 
  ~~~~~
    \begin{minipage}[h]{0.45\linewidth}
    \center{\includegraphics[width=1\linewidth]{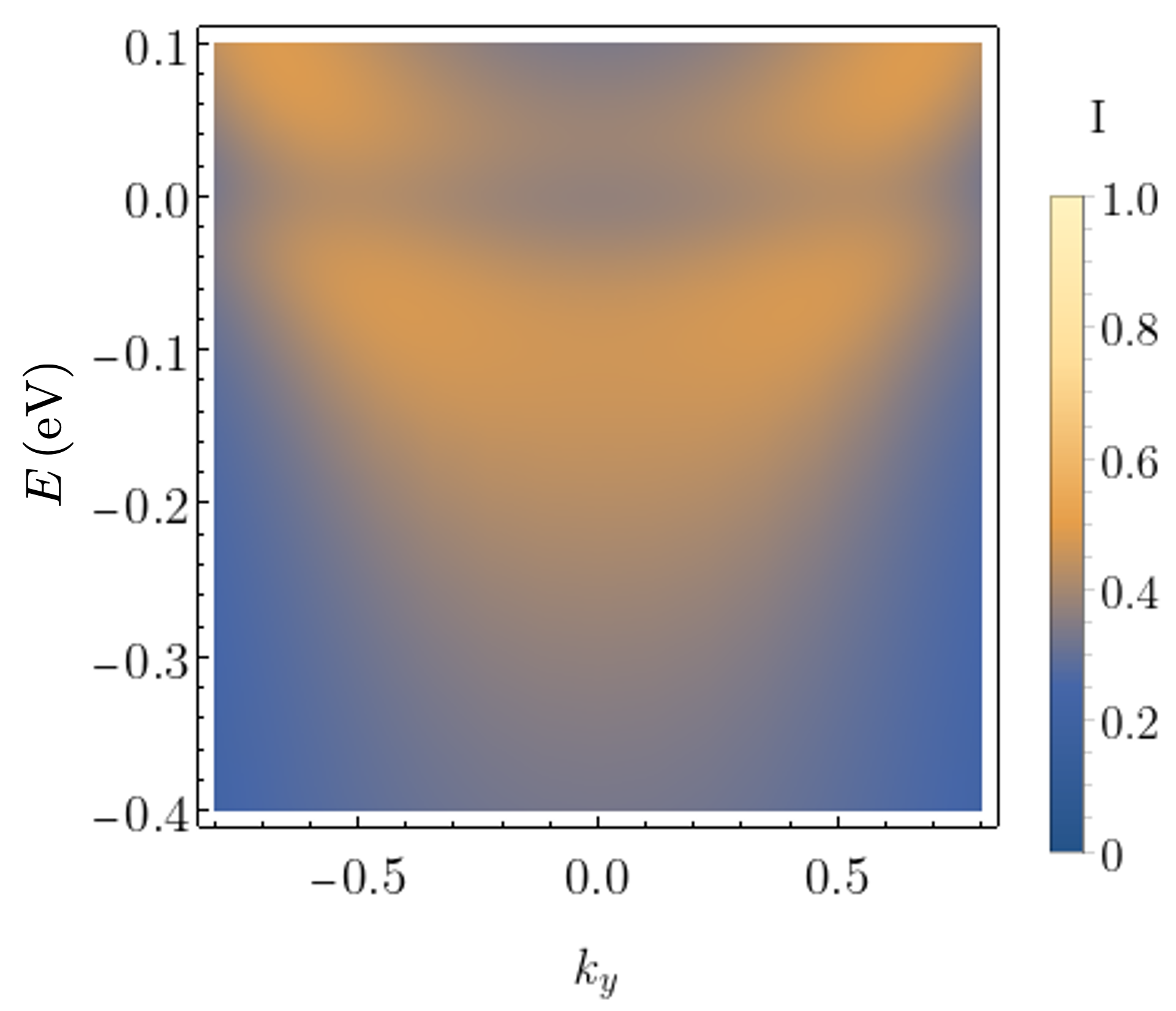}}
     \\b)
    \end{minipage} 
\caption{(a) Fermi surface of unhybridized and hybridized bands in the underoped regime with 2 CuO layers. The red line shows physical layer bands ($\epsilon_c(k)\pm e_\perp(k)=0$), black dashed line shows the hidden first layer band
($\epsilon_f(k)=0$), while blue lines are hybridized dispersions ($E_{\pm}(k)=0$).
(b) Theoretical computations of the $c$ spectral density in the antinodal region for $T=150$K with the MFL self energy in Eq.~(\ref{eqn:marginal_self_energy}) with $\lambda = 0.5$ and other parameters from Fig.~\ref{fig:chen_fl_cut}; compare these to Figs.~1,2 in Ref.~\cite{chen2019incoherent}.}
\label{fig:antinode3}
\end{figure}
The resulting Fermi surfaces, shown in Fig.~\ref{fig:antinode3}(a), are similar to that obtained for Bi2201: upon hybridization, the two large Fermi surfaces (red lines) split by the bilayer coupling, transform into two hole pocket Fermi surfaces (blue lines).  As previously discussed,  most of the spectral weight of $A_{cc} ({\bf k},\omega=0)$ is confined to sides of the hole pockets facing the $\Gamma$ point, implying the side facing the $(\pi,\pi)$-point will not be visible in ARPES experiments, thus explaining the emergence of Fermi arcs. In Fig.~\ref{fig:antinode3}(b), we plot the spectral function in the antinodal region, which is significantly smeared due to the MFL form of the self-energy, and exhibits a hybridization gap around zero energy.

It is instructive to consider the evolution of the EDCs and MDCs between the overdoped (Fermi-liquid) and underdoped (non-Fermi-liquid) regimes, which we present in Fig.~\ref{fig:chen_experiment}.  
To further illustrate this point, we made EDC and MDC cuts in the antinodal region and compared them to the Fermi liquid results in Fig.~\ref{fig:chen_experiment}(a),(b).
While the EDC exhibits a transfer of the spectral weight from low to higher energies, between the overdoped and underdoped regime (arising from the emergence of a gap, as shown in Fig.~\ref{fig:antinode3}(b)), the corresponding MDCs exhibit only weak changes. These results are in good agreement with the experimental observations shown in Fig.~\ref{fig:chen_experiment}(c),(d).

%

Next, we employ the results of Section~\ref{sec:frac} on the influence of a SYK spin liquid on the second hidden layer to improve upon the MFL self energy shown in Eq.~(\ref{eqn:marginal_self_energy}).
The frequency dependence of the real and imaginary parts of the self energies $\Sigma_c (\omega)$ and $\Sigma_f (\omega)$ around the Fermi energy are shown in Fig.~\ref{fig:SelfEn}.
Using these self energies, we compute the spectral function in the underdoped regime at a temperature of $T=150$K, and present the EDC and MDC curves in Fig.~\ref{fig:chen_experiment} (c),(d).
These results show better agreement with the experimental observations in Fig.~\ref{fig:chen_experiment} (e),(f) than the ones previously shown using the MFL form of the self-energy shown in Eq.~(\ref{eqn:marginal_self_energy}). In particular, the EDC curve is significantly flatter than in the marginal Fermi liquid case.  

Further improvements on the electron self energy can be made by choosing a realistic spin liquid of a short-range Hamiltonian in the second hidden layer. An attractive candidate is a gapless $\mathbb{Z}_2$ spin liquid which has attracted significant recent interest \cite{TSMPAF99,WenPSG,Becca01,SenthilIvanov,SenthilLee05,Kitaev2006,Becca13,Sandvik18,Becca18,Becca20,Imada20,Gu20,shackleton2021,Gu2021}, but the paramagnon fractionalization approach allows for a very wide range of possibilities on the second hidden layer. 
In Appendix~\ref{app:z2} we compute gapless $\mathbb{Z}_2$ spin liquid self energies of the $c$ and $f$ fermions, and the EDC result is shown in Fig.~\ref{fig:EDC_theory_SL}. We see that even at small broadening, the EDC curve in the underdoped regime have a similar frequency dependence to Fig.~\ref{fig:chen_experiment}.

\begin{figure}
\begin{minipage}[h]{0.4\linewidth}
  \center{\includegraphics[width=1\linewidth]{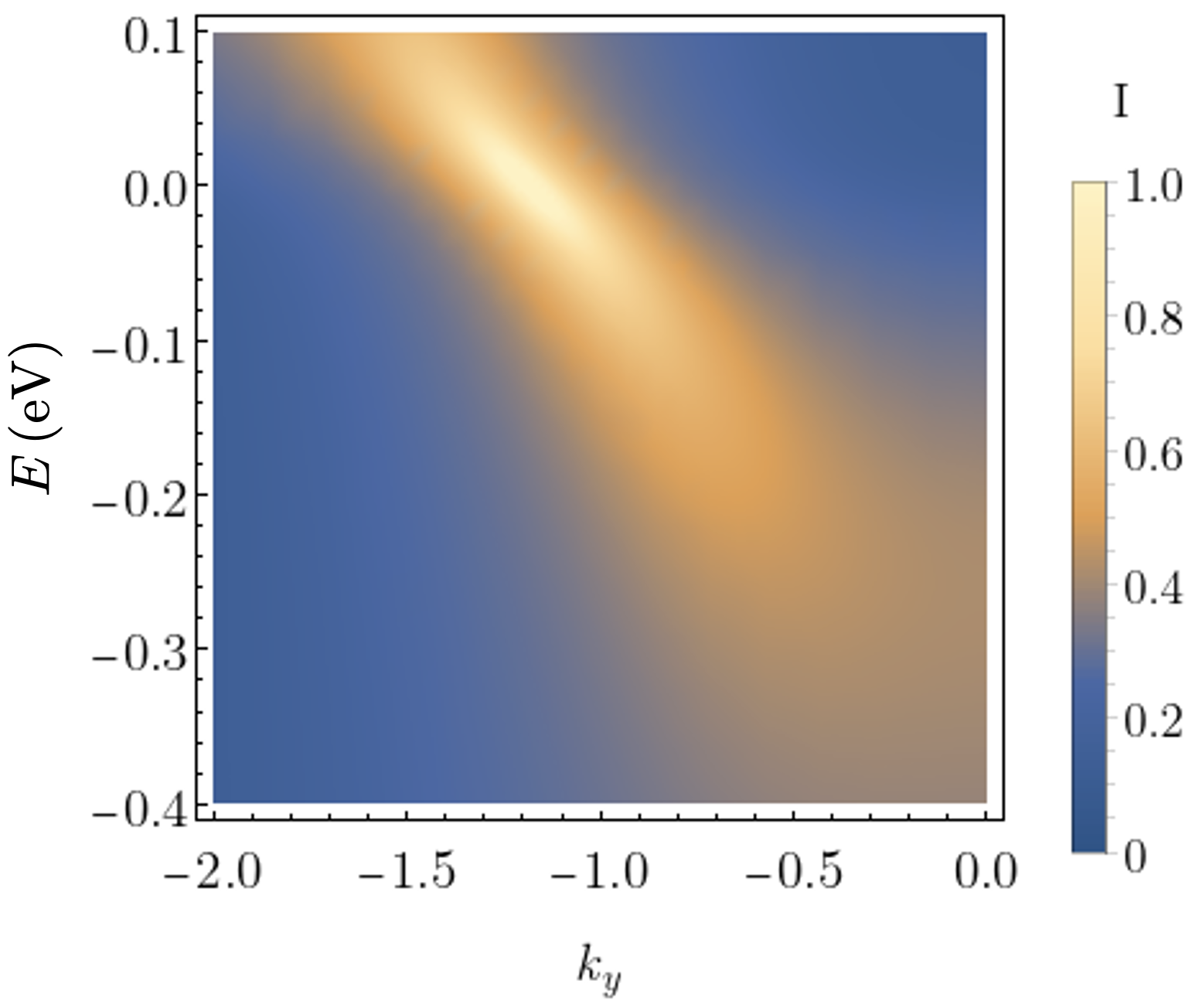}}
 \\overdoped phase.
  \end{minipage} 
  ~~~~~   
    \begin{minipage}[h]{0.4\linewidth}
    \center{\includegraphics[width=1\linewidth]{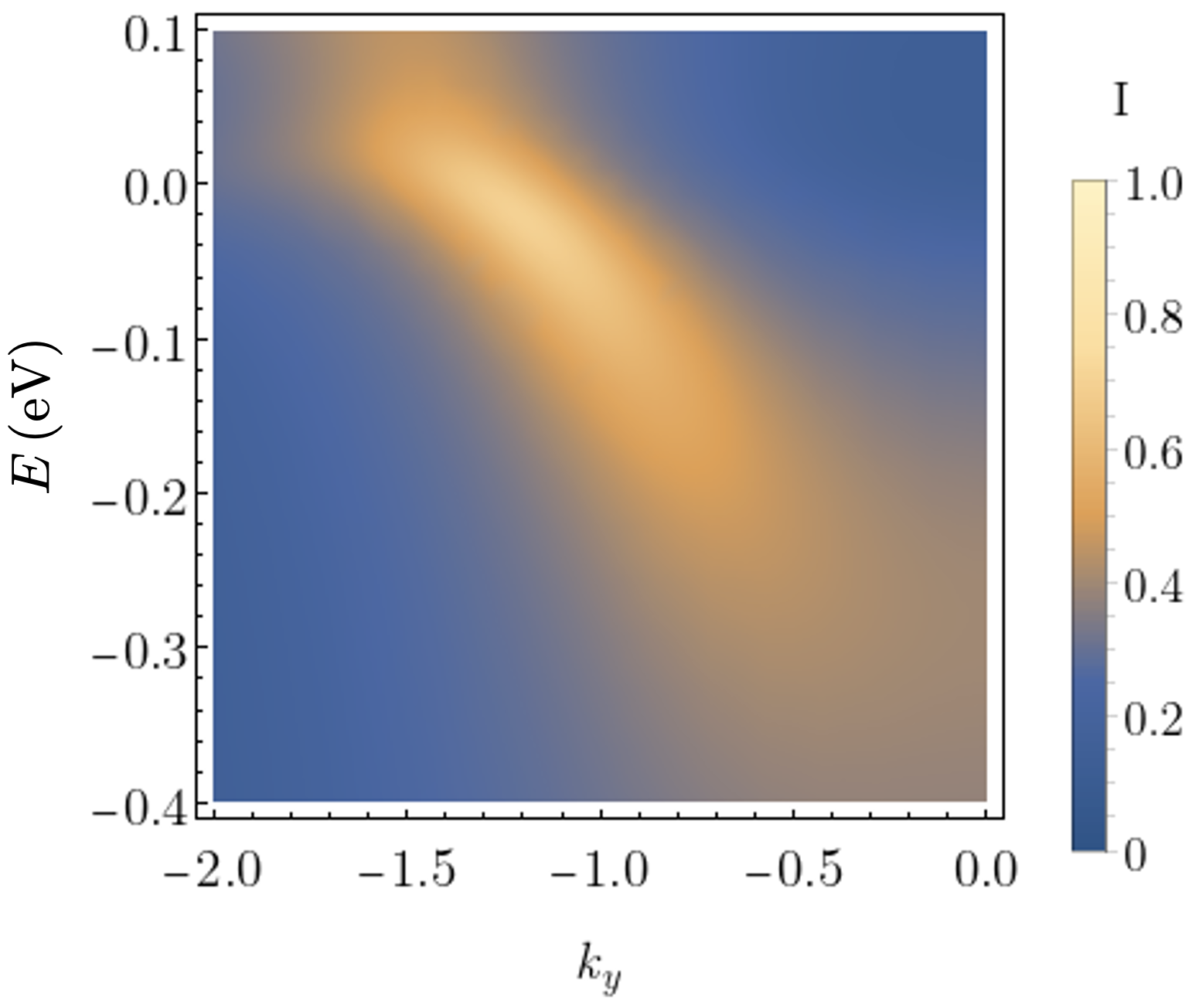}}
     \\underdoped phase.
    \end{minipage} 
\caption{Computations of ARPES intensity for $T=150$K and other parameters from Fig.~\ref{fig:chen_fl_cut} at the nodal cut: $k_x=-1.15$. Compare to Figs.~1G,H in Chen {\it et al.\/} \cite{chen2019incoherent}.}
\label{fig:node2}
\end{figure}

Finally, we consider the transition from the Fermi liquid to the pseudogap phase in the nodal region. A comparison of the theoretical spectral functions near the nodal points  (see Fig.~\ref{fig:node2}) shows a very similar structure in the overdoped and underdoped regimes, in agreement with the experimental findings \cite{chen2019incoherent}. This is contrast to the behavior near the antinodal points, where the underdoped and overdoped regimes are very different.


\section{Conclusions}

This paper has compared experimental observations with the `ancilla' theory of the pseudogap metal in the Refs.~\cite{Zhang2020,Zhang2020_1,nikolaenko2021}, which we have now dubbed the theory of `paramagnon fractionalization'.

In Section~\ref{sec:2201} we examined the dispersion of electronic excitations in the pseudogap metal, as summarized in the paradigmatic observations of He {\it et al.\/} \cite{He2011} in Fig.~\ref{fig:He}(e),(f). 
At the anti-node the electronic dispersion has a minimum gap at wavevectors $k_{G1}, k_{G2}$ which differ from the Fermi wavevectors ($k_{F1}$, $k_{F2}$) found at temperatures above the pseudogap onset. We showed that this feature could be reproduced in the paramagnon fractionalization model, as shown in Figs.~\ref{fig:He}(a),(c) (similar features are also present in the pair density wave state \cite{Lee2014}, as we review in Appendix~\ref{app:sdwpdw}). 
In the nodal region of the Brillouin zone, there is little difference in the observed electronic dispersion between the underdoped and overdoped regimes, as shown in Fig.~\ref{fig:He}(f), which is also what we find in the paramagnon fractionalization model (see Fig.~\ref{fig:He}(b,d)) (in contrast, the PDW model does modify the underdoped nodal spectrum, as discussed in Appendix~\ref{app:sdwpdw}).
For the same parameters used for the dispersions in Figs.~\ref{fig:He}(a)-(d), our model also yields a reasonable shape of the pocket Fermi surfaces, with suppressed spectral density at the `back side' of the pockets, as shown in Fig.~\ref{fig:brillouin_zone}. The ability to fit this wide range of data in both the nodal and anti-nodal regions, with the only fitting parameters being the hopping parameters in (\ref{eq:ef}) (the Higgs condensate $\phi$ is strongly constrained by the size of the pseudogap energy) is a particular strength of our theory. 

In Section~\ref{sec:2212} we examined the evolution of the MDC and EDC curves in the anti-nodal region between the underdoped and overdoped metals, as observed in Bi2212 by Chen {\it et al.} \cite{chen2019incoherent}. For this, it was necessary to include a frequency-dependent electronic self energy. We first modeled this by the phenomenological marginal Fermi liquid form \cite{Varma1989} in Eq.~(\ref{eqn:marginal_self_energy}), and found
that this provided a reasonable description of these observations.

In Section~\ref{sec:frac}, we proposed that the primary contribution to the frequency dependence of the electronic self energy arose from an exchange coupling to gapless fractionalized spinons in the second layer of hidden spins produced by fractionalizing the paramagnon; all other fluctuations are gapped out in this theory of pseudogap metal \cite{Zhang2020,Zhang2020_1}, apart from those associated with electronic Fermi surface. A marginal Fermi liquid self energy is obtained in such a model if we assume that the spinons are described by a SYK spin liquid \cite{SY,Parcollet1,CGPS}, and assume the electronic local density of states is energy independent. Section~\ref{sec:frac} improved on this model by including the energy dependence of the electronic local density of states which is computed in 
Section~\ref{sec:2212}. 
The resulting spectral functions in Fig.~\ref{fig:chen_experiment}(c),(d) improved the agreement with the experimental observations in Fig.~\ref{fig:chen_experiment}. 

One promising description of the spinons in the second hidden layer in a model without disorder is the gapless $\mathbb{Z}_2$ spin liquid, which has received significant attention recently \cite{TSMPAF99,WenPSG,Becca01,SenthilIvanov,SenthilLee05,Kitaev2006,Becca13,Sandvik18,Becca18,Becca20,Imada20,Gu20,shackleton2021,Gu2021}. Appendix~\ref{app:z2} presented results for the electronic self energy using this gapless $\mathbb{Z}_2$ spin liquid, and showed that this liquid can yield significant damping of the fermions at the pseudogap energy in the anti-nodal region, and also improve correspondence with observations. Other spin liquids can also fit into the paramagnon fractionalization theory, and used to compute the electron self energy as in Appendix~\ref{app:z2}.

In our view, given the simplicity of our model, the natural description it provides of numerous observations is encouraging. 
A more conclusive test of our model can be provided by a more direct detection of a distinctive feature of our theory: the fractionalized spinon excitations present in the second hidden layer of spins created by paramagnon fractionalization. Such excitations are not present in the heavy Fermi liquid of the Kondo lattice, and are required by the existence of a non-Luttinger volume Fermi surface in a single band model \cite{FLSPRL,TSFL04}.
The spinons can be detected by a clearer signal of their consequences in the electronic spectrum, or in direct measurements of the spin spectrum by neutron scattering or other probes. There is evidence for spinons in the undoped antiferromagnet \cite{Piazza15}, and it would be interesting to study its evolution under non-zero doping.

\subsection*{Acknowledgements}

We are grateful to Sudi Chen, Makoto Hashimoto, Yu He, Zhi-Xun Shen, and Kejun Xu for very helpful discussions on their ARPES observations. We also thank Weijiong Chen, Seamus Davis, and Shuqiu Wang for valuable comments.
The research of A.N. and S.S. on the trial wavefunction of the Hubbard model, which applies also to optical lattices of ultracold fermionic atoms, was supported by  the U.S. Department of Energy under Grant DE-SC0019030. The research of M.T., S.S., and Y.Z. on spin liquids was supported by the U.S. National Science Foundation grant No. DMR-2002850, by the Gordon and Betty Moore Foundation’s EPiQS Initiative Grant GBMF8683, and by the Simons Collaboration on Ultra-Quantum Matter which is a grant from the Simons Foundation (651440, S.S.). The research of E.M. and D.K.M. was supported by the U. S. Department of Energy, Office of Science, Basic Energy Sciences, under Award No. DE-FG02-05ER46225


\appendix
\section{Gapless $\mathbb{Z}_2$ spin liquid on the second hidden layer}
\label{app:z2}

We now discuss the behavior of the spectral function on the physical $c$-electron layer when the system on the second hidden layer is considered to be a gapless $\mathbb{Z}_2$ spin liquid \cite{TSMPAF99,WenPSG,Becca01,SenthilIvanov,SenthilLee05,Kitaev2006,Becca13,Sandvik18,Becca18,Becca20,Imada20,Gu20,shackleton2021,Gu2021}. While the first two layers are coupled via $J_K$ coupling constant, the second and the third layers are perturbatively coupled via the $J_\perp$ and $\widetilde{J}_K$ exchange interactions (see Fig.~\ref{fig:layers}). 

\begin{figure}
\includegraphics[width=.49\linewidth]{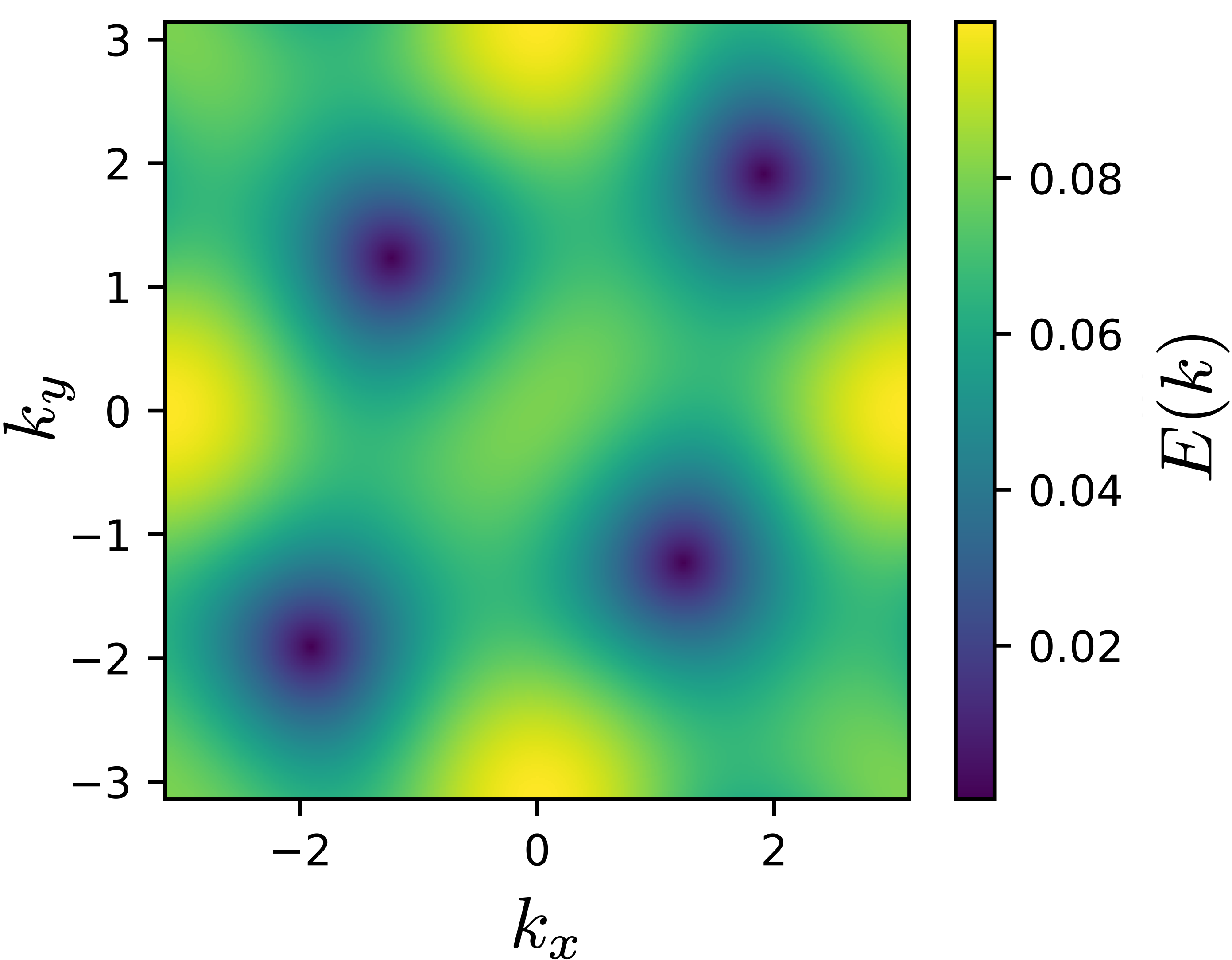}
\caption{Spin liquid dispersion  \eqref{eq:dispersion} with parameters $\chi = 0.02$, $\eta = 0.025$ and $\gamma_1 = 0.0075$}
\label{fig:dispersion}
\end{figure}

We consider a gapless $\mathbb{Z}_2$ spin liquid Hamiltonian \cite{shackleton2021} in the following form 
\beq
H = \sum_k \begin{pmatrix}
\widetilde{f}_{k\uparrow}^\dagger & \widetilde{f}_{-k\downarrow}
\end{pmatrix}
\begin{pmatrix}
0 & \Delta_k\\
\bar{\Delta}_k & 0
\end{pmatrix}
\begin{pmatrix}
\widetilde{f}_{k\uparrow} \\ \widetilde{f}_{-k\downarrow}^\dagger
\end{pmatrix} = \sum_k \psi_k^\dagger 
\begin{pmatrix}
0 & \Delta_{1k}-i\Delta_{2k}\\
\Delta_{1k}+i\Delta_{2k} & 0
\end{pmatrix}\psi_k
\eeq
where $\psi_k$ is a Nambu spinor, $\Delta_{1k}$ and $\Delta_{2k}$ are the momentum dependent functions that have the following form
\beq\label{eq:delta}
&&\Delta_{1k} = 2\chi(\cos{k_x}+\cos{k_y})+4\gamma_1\sin{k_x}\sin{k_y},\\
&&\Delta_{2k} = 2\eta(\cos{k_x}-\cos{k_y}).
\eeq
The dispersion is
\beq\label{eq:dispersion}
|\Delta_k|^2= [2\chi(\cos{k_x}+\cos{k_y})+4\gamma_1\sin{k_x}\sin{k_y}]^2+[2\eta(\cos{k_x}-\cos{k_y})]^2.
\eeq
and $E_k = |\Delta_k|$. Behavior of the spin liquid dispersion at fixed $\chi$, $\eta$ and $\gamma_1$ is shown in Fig.~\ref{fig:dispersion}.  
Using the isospin matrices
\beq
\tau = \{\tau_1,\tau_2,\tau_3\} = 
\left\{
\begin{pmatrix}
0 & 1\\
1 & 0
\end{pmatrix},
\begin{pmatrix}
0 & -i\\
i & 0
\end{pmatrix},\begin{pmatrix}
1 & 0\\
0 & -1
\end{pmatrix}\right\}
\eeq
we can rewrite the Hamiltonian in the simpler form and obtain
\beq
H = \sum_k \psi_k^\dagger [\Delta_{1k}\tau_1+\Delta_{2k}\tau_2]\psi_k = \sum_k \psi_k^\dagger(h_k\cdot\tau)\psi_k,
\eeq
where $h_k=(\Delta_{1k},\Delta_{2k},0)$.

We now turn to computing the spin-spin correlation function using the Nambu-Gorkov Green's function 
\beq
G_{\alpha\beta}(k,\tau) = -\langle T\psi_{k\alpha}(\tau)\psi_{k\beta}^\dagger(0)\rangle = 
-\begin{pmatrix}
\langle T \widetilde{f}_{k\uparrow}(\tau)\widetilde{f}_{k\uparrow}^\dagger(0)\rangle & \langle T\widetilde{f}_{k\uparrow}(\tau)\widetilde{f}_{-k\downarrow}(0)\rangle\\
\langle T\widetilde{f}_{-k\downarrow}^\dagger(\tau)\widetilde{f}_{k\uparrow}^\dagger(0)\rangle & \langle T\widetilde{f}_{-k\downarrow}^\dagger(\tau)\widetilde{f}_{-k\downarrow}(0)\rangle
\end{pmatrix}
\eeq
Performing the Fourier transform and using the expression for the bare Green's function
\beq
G(k,i\omega_n) = \frac{1}{i\omega_n - h_k},
\eeq
we obtain the Green's function in the following form
\beq
G(k,i\omega_n) = \frac{1}{(i\omega_n)^2 - E_k^2} 
\begin{pmatrix}
i\omega_n & \Delta_k\\
\bar{\Delta}_k & i\omega_n
\end{pmatrix}.
\eeq
where the dispersion is given by $E_k = |\Delta_k|$.
We now compute the spin susceptibility 
\beq
\chi_{ab}(i\nu_n,q) = \delta_{ab}\chi_{aa}(i\nu_n,q) =  \delta_{ab}\int_0^\beta d\tau\langle T S_a(\tau,q)S_a(0,-q)\rangle e^{i\nu_n \tau}
\eeq
We will use $S_3$ component for simplicity which in Nambu basis reads
\beq
S_3(q) = \sum_{k} \psi_{\alpha, k+q}^\dagger I_{\alpha\beta} \psi_{\beta, k} = \sum_{k} \psi_{k+q}^\dagger \cdot \psi_{k}
\eeq
where $I_{\alpha\beta}$ is the identity matrix. Dropping the subscripts, expression for the spin susceptibility reads
\beq
\chi(i\omega_n, q) = -\frac2\beta\sum_{k,\nu_m}  \left[\frac{i\nu_m(i\nu_m+i\omega_n) + \Delta_{1,k}\Delta_{1,k+q}+\Delta_{2,k}\Delta_{2,k+q}}{((i\nu_m+i\omega_n)^2-E_{k+q}^2)((i\nu_m)^2-E_{k}^2)}\right]
\eeq
Performing the summation over Matsubara frequencies, we obtain
\beq\nonumber
&&\chi(i\omega_n,q) = \frac{1}{2}\sum_{k} w^+_{k, k+q} \left\{\frac{1}{i\omega_n - E_{k+q}+E_k}-\frac{1}{i\omega_n + E_{k+q}-E_k}\right\}(n_F(E_{k+q})-n_F(E_k)) \\
&&-\frac{1}{2}\sum_{k} w^-_{k, k+q} \left\{\frac{1}{i\omega_n - E_{k+q}-E_k}-\frac{1}{i\omega_n + E_{k+q}+E_k}\right\}(1-n_F(E_{k+q})-n_F(E_k)),
\eeq
where the coherence factor reads
\beq
w^\pm_{k, k+q}= 1\pm\frac{\Delta_{1,k}\Delta_{1,k+q} + \Delta_{2,k}\Delta_{2,k+q} }{E_{k}E_{k+q}}.
\eeq
We will present results for self energy and ARPES at zero temperature in this section. At zero temperature the spectral function for the spin susceptibility is simply
\beq
\rho_\chi(\omega,q) = -\frac{1}{2}\sum_{k} w^-_{k, k+q} \left(\delta(\omega - E_{k+q} - E_k) -\delta(\omega + E_{k+q} + E_k)\right) 
\eeq

\begin{figure}
\includegraphics[width=.48\linewidth]{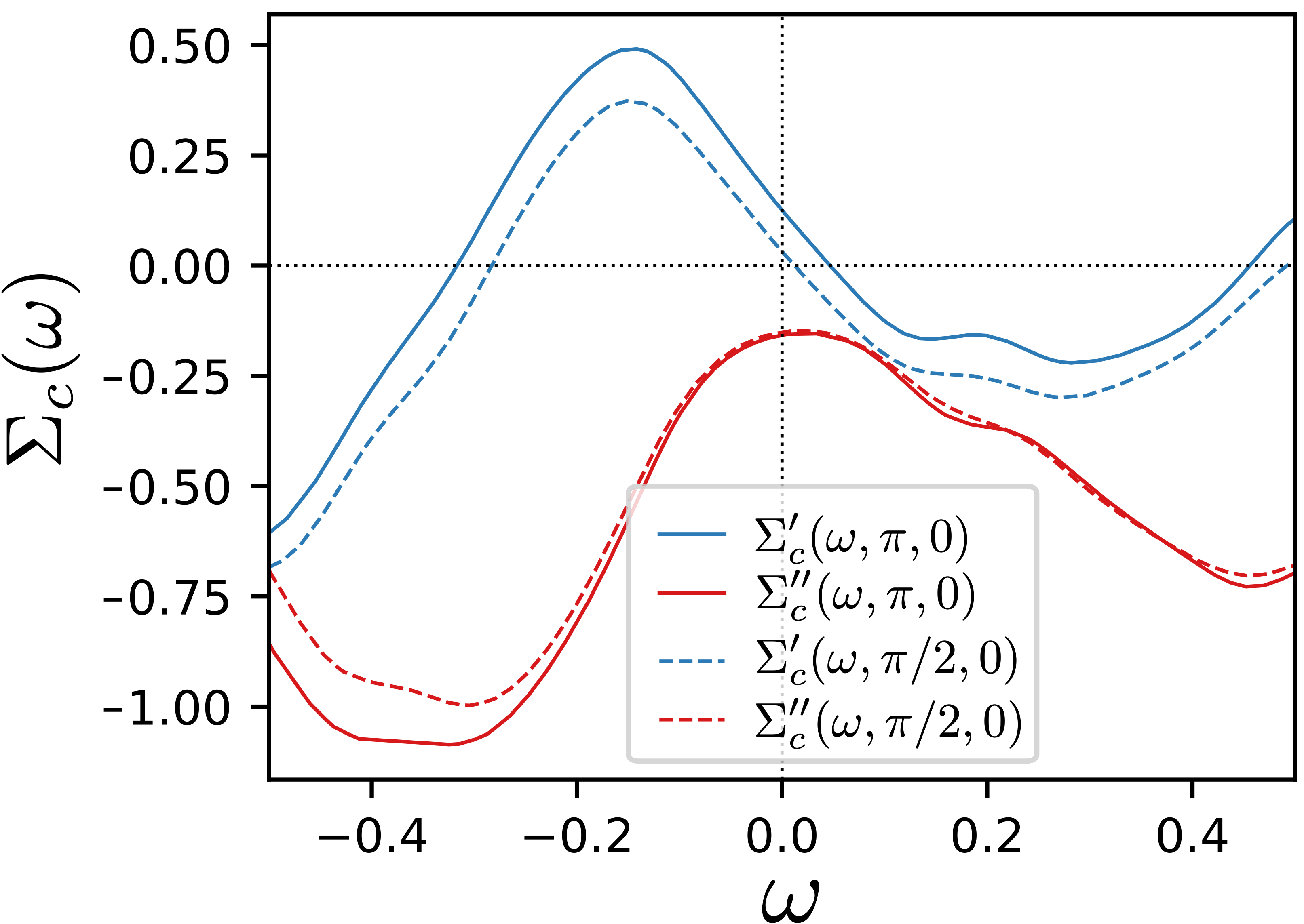}\,\,
\includegraphics[width=.49\linewidth]{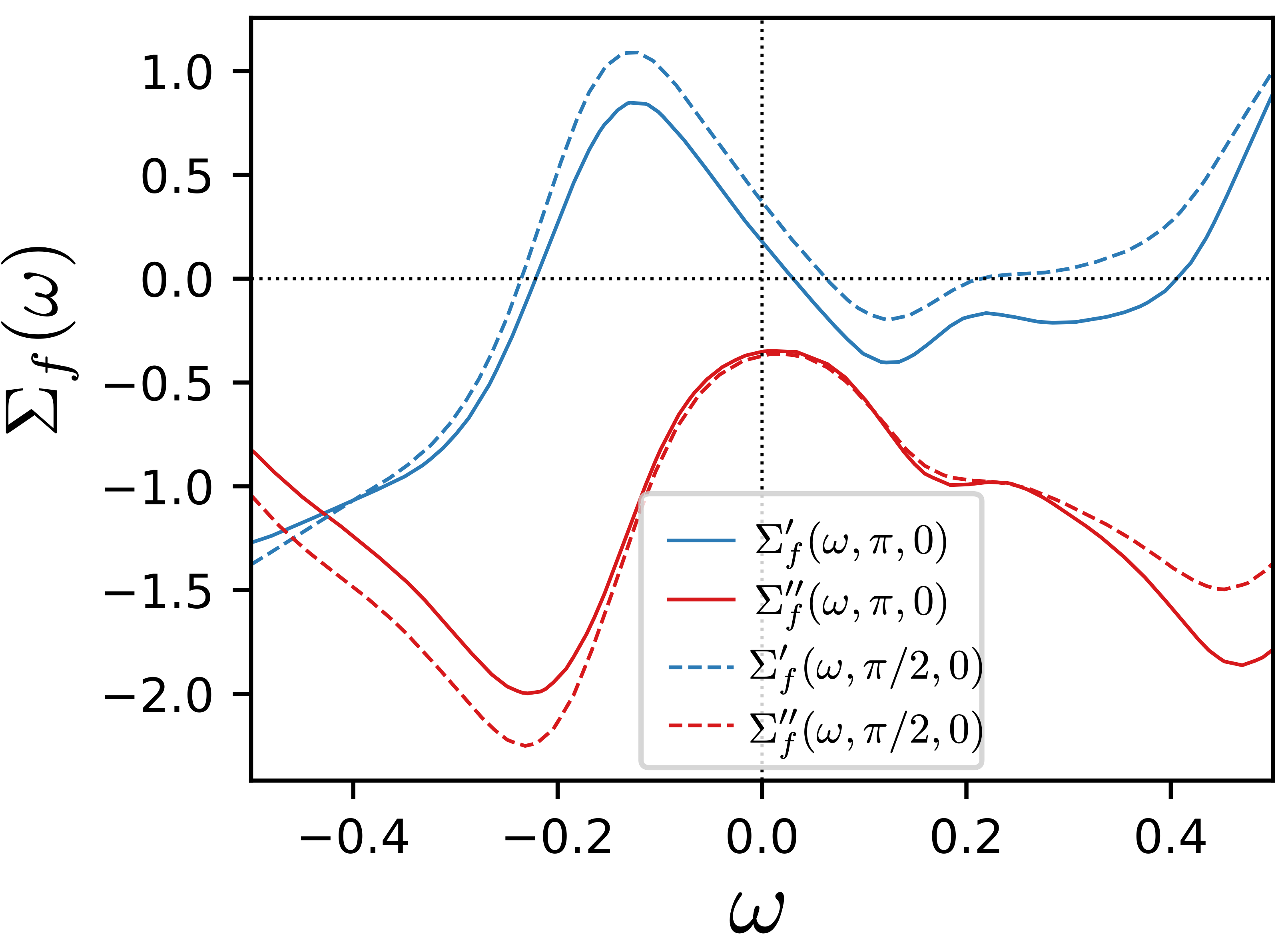}
\caption{Real (blue lines) and imaginary (red lines) parts of self energies in the case of gapless $\mathbb{Z}_2$ spin liquid for $c$ electrons (left) and $f$ fermions (right) with $\tilde{J}_K^2 =J_\perp^2= 1$. We denote real part of a self energy as $\Sigma'(\omega,k_x,k_y)$ and imaginary part as $\Sigma''(\omega,k_x,k_y)$. Dashed lines are computed in nodal region at point $(k_x,k_y) = (\pi/2,0)$ and the solid lines are in the antinodal region at momentum $(\pi,0)$. The spin liquid parameters:  $\chi = 0.02$, $\eta = 0.025$ and $\gamma_1 = 0.0075$.}
\label{fig:SelfEnSLChi002}
\end{figure}

\begin{figure}
\begin{minipage}[h]{1\linewidth}
  \center{\includegraphics[width=.5\linewidth]{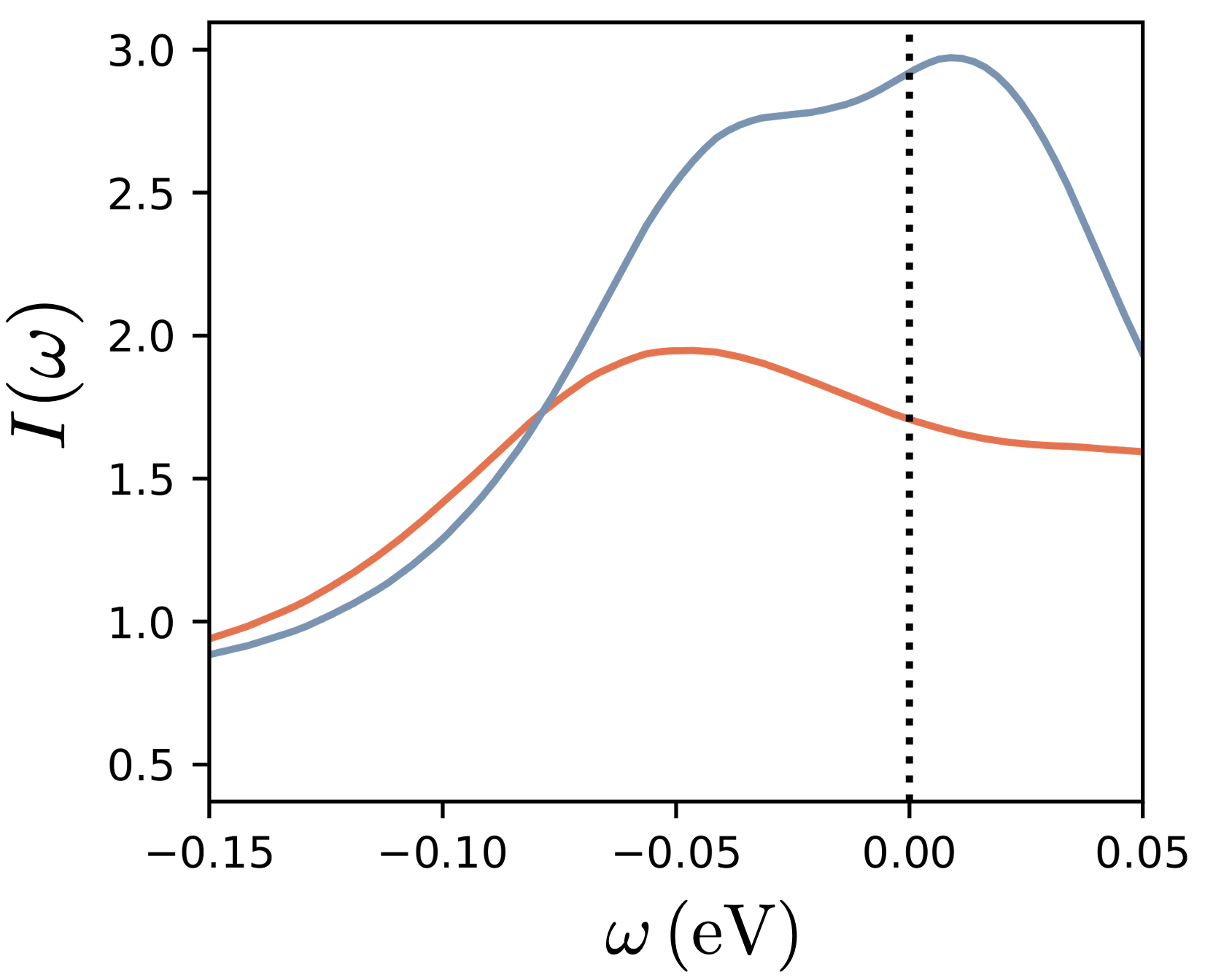}}
  \end{minipage}
\caption{ARPES EDC curve in the antinodal region at zero temperature with spin liquid self energy \eqref{SL_SE} with $\tilde{J}_K = 0.25$ and $J_c = 0.85$. Impurity induced broadening for the $c$ - electron layer is $\Gamma_c = 0.03$ and for the $f$-fermion layer is $\Gamma_f = 0.1$. The blue line respresents the overdoped case, while orange line correspond to an underdoped case when the physical layers are coupled to the hidden layers with $\phi=0.15$. Spin liquid dispersion parameters are $\chi = 0.02$, $\eta = 0.025$ and $\gamma_1 = 0.0075$. }
\label{fig:EDC_theory_SL}
\end{figure}

We can now find self energy similarly as in Section \ref{sec:frac}. At zero temperature  $f$ fermion self energy reads
\beq
\Sigma_f(k,i\omega_n) = J_\perp^2 \sum_{q,q'} \int d\omega \rho_f(k-q,\omega) w^-_{q,q+q'} \left(\frac{\theta(\omega) }{i\omega_n - \omega -E_{q+q'} - E_q} - \frac{\theta(\omega) - 1}{i\omega_n - \omega +E_{q+q'} + E_q}\right)\nonumber
\eeq
with the hybridized $f$ fermion spectral function 
\beq
\rho_{f}(\omega,k) =   v_k^2 \delta(\omega - \xi^+_k) + u_k^2 \delta(\omega - \xi^-_k).
\eeq
where $v_k$, $u_k$ and $\xi_k^\pm$ are given by \eqref{ukvk}-\eqref{xik}.
We finally obtain the expression for the self energy 
\beq
\Sigma_f(k,i\omega_n) &&= J_\perp^2\int \frac{d^2q\,d^2q'}{(2\pi)^4}  w^-_{q,q+q'}\Bigg[v^2_{k-q} \left(\frac{\theta(\xi^+_{k-q}) }{i\omega_n - \xi^+_{k-q} -E_{q+q'} - E_q} + \frac{\theta(-\xi^+_{k-q}) }{i\omega_n - \xi^+_{k-q} 
+E_{q+q'} + E_q}\right)\nonumber \\ &&+u_{k-q}^2\left(\frac{\theta(\xi^-_{k-q}) }{i\omega_n - \xi^-_{k-q} -E_{q+q'} - E_q} + \frac{\theta(-\xi^-_{k-q})}{i\omega_n - \xi^-_{k-q} +E_{q+q'} + E_q}\right)\Bigg].\label{SL_SE}
\eeq
We can now analytically continue and compute real and imaginary parts in a similar way as we did in Section \ref{sec:frac}. Self energy for the $c$ electron layer can be computed similarly with the corresponding spectral function $\rho_c(\omega)$. 
We show behavior of the self energies for both $c$ and $f$ fermion layers in Fig.~\ref{fig:SelfEnSLChi002}. 
In comparison to Fig.~\ref{fig:SelfEn} from the SYK model, we note more structure in the $\omega$ dependence, with small values near the Fermi energy, and larger values near the pseudogap energy. We show behavior of the spectral function for Bi2212 case in Fig.~\ref{fig:EDC_theory_SL} and find that due to non trivial structure of the self energy, we are able to adjust parameters such that the model with $\mathbb{Z}_2$ spin liquid self energy fit the experimental data in Fig.~\ref{fig:chen_experiment} (e) even better. 

\section{Electronic structure in the SDW and PDW phases}
\label{app:sdwpdw}

This appendix briefly recalls the form of the electronic structure in the SDW and PDW phases, two other candidate models for the pseudo-gap region, and compares them with the paramagnon fractionalization model.

\subsection{Spin density wave}
\label{sec:sdw}

\begin{figure}
\begin{minipage}[h]{0.33\linewidth}
  \center{\includegraphics[width=1\linewidth]{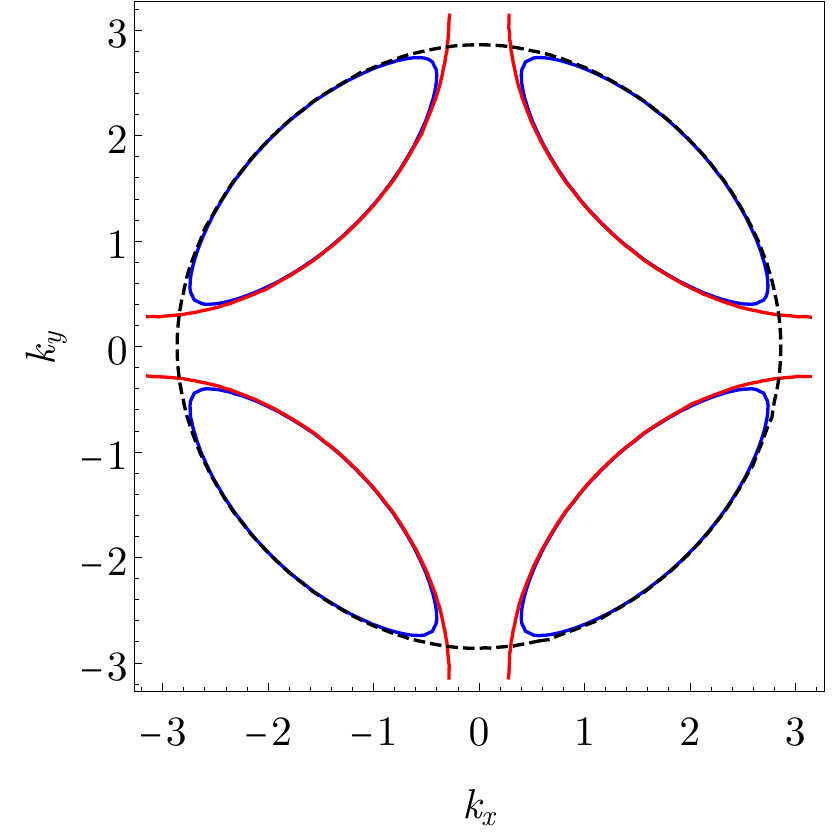}}
 \\a) Full Brillouin zone.
  \end{minipage} 
  ~~~~~
    \begin{minipage}[h]{0.33\linewidth}
    \center{\includegraphics[width=1\linewidth]{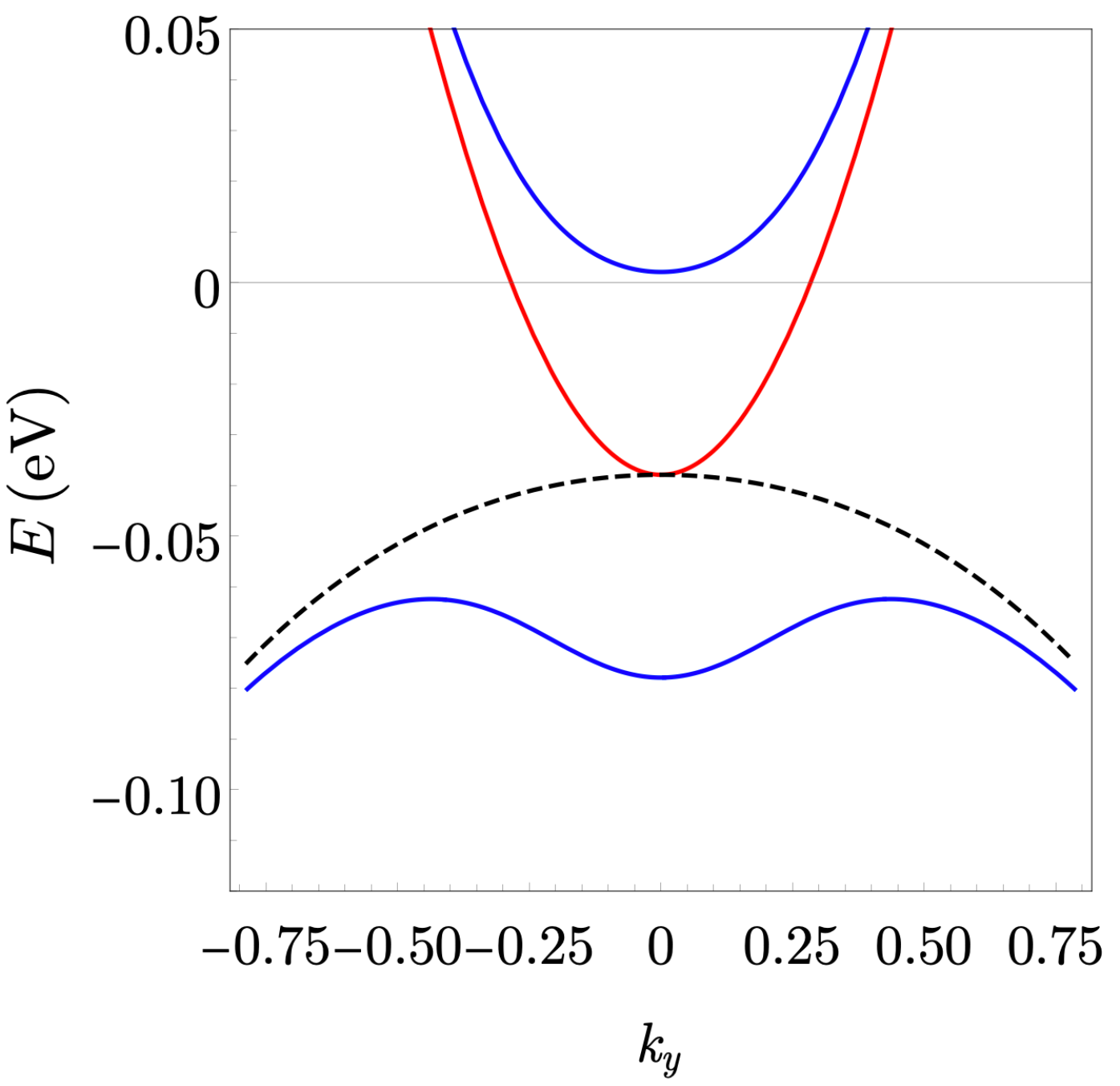}}
     \\b) $k_x=-\pi$.
    \end{minipage} 
\caption{a) Fermi surfaces of SDW bands. The red line shows the physical layer band, black dashed line shows shifted band while blue lines are hybridized bands.
(b) Bands in the antinodal region display Bogoliubov-like dispersion. The tight-binding parameters are same as in Fig.~\ref{fig:brillouin_zone}. }
\label{fig:sdw}
\end{figure}

While there is currently no experimental evidence for the existence of a long-ranged spin-density-wave (SDW) phase in the pseudo-gap regime, it is instructive to compare its electronic structure with that obtained from the paramagnon fractionalization model. Specifically, a SDW phase with ordering wave-vector $\vec{Q}=(\pi, \pi)$ yields an electronic structure that is equivalent to a paramagnon fractionalization model with the band dispersion $\epsilon_f(\vec{k})=\epsilon_c(\vec{k}+\vec{Q})$ and $\phi \rightarrow m$, where $m$ is the SDW order parameter. The resulting Fermi surface in the SDW phase, and band structure near $k_x=\pi$ with $m=0.04eV$ are shown in Figs.~\ref{fig:sdw}(a) and (b), respectively.
At first sight, the hybridized SDW bands look similar to those of the paramagnon fractionalization model. However, contrary to the paramagnon fractionalization model, the shifted physical layer band, i.e., $\epsilon_c(\vec{k}+\vec{Q})$ [the dashed line in Fig.~\ref{fig:sdw}(b)] lies below the Fermi level, and as a result, the minimum energy of the hybridized upper band occurs at $k_y=0$. Though ARPES experiments (at $T=0$) cannot probe the band dispersion above the Fermi energy, the difference in the upper hybridized bands can be seen in the MDC (momentum distribution curve) cuts at zero energy. The MDC cut for an SDW metal would have a maximum at $k_y=0$, in contrast to the paramagnon fractionalization model. That contradicts the experiments in the similar material Bi2212\cite{chen2019incoherent}, where the MDC cuts do not change significantly by going from Fermi liquid to a pseudogap phase. We will show this in Section~\ref{sec:2212} on ARPES in Bi2212.

\subsection{Pair density wave}
\label{sec:pdw}

\begin{figure}
\begin{minipage}[h]{0.4\linewidth}
  \center{\includegraphics[width=1\linewidth]{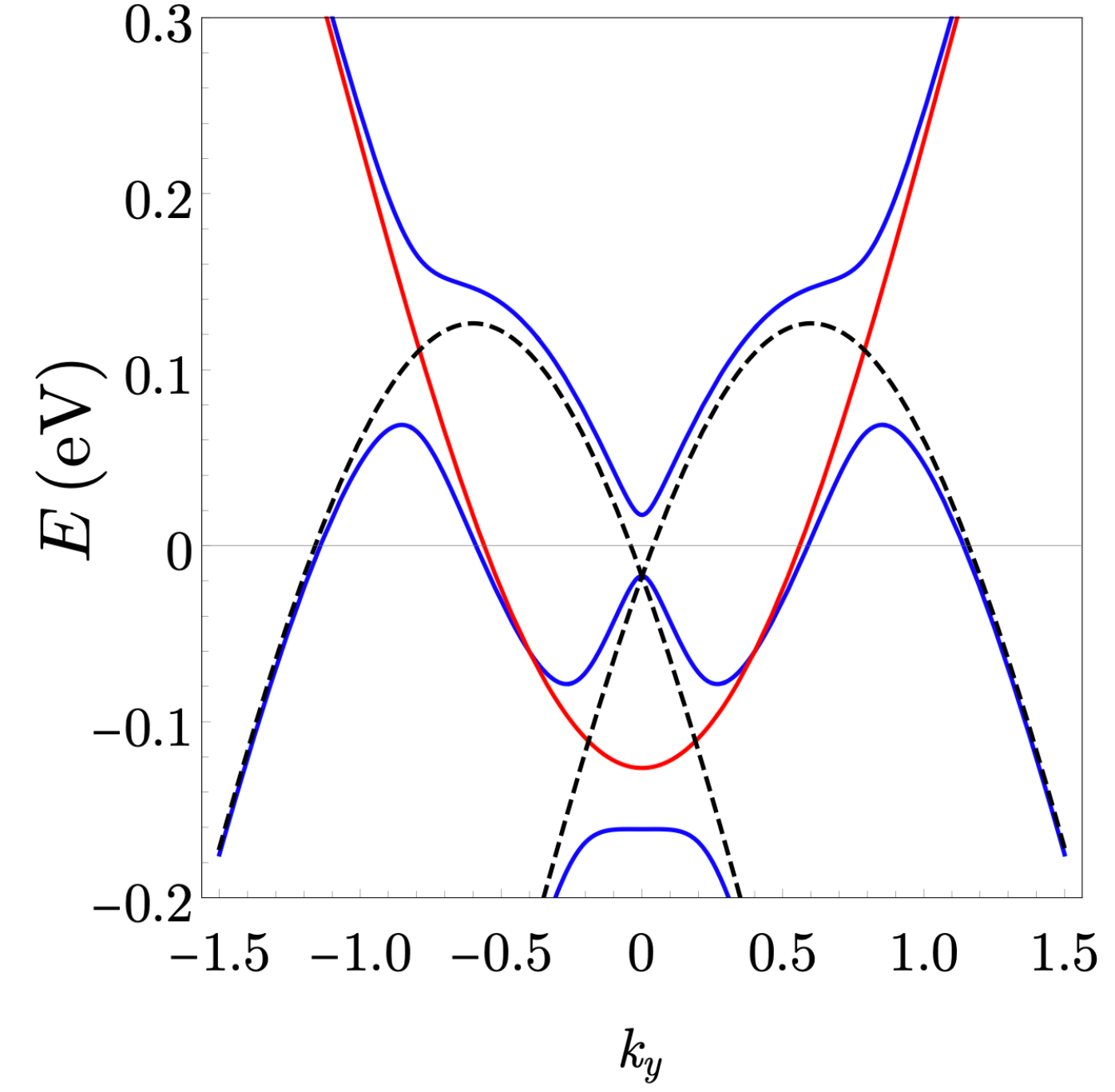}}
 \\a) Band structure in the nodal region region $k_x=-2$.
  \end{minipage} 
  ~~~~~
    \begin{minipage}[h]{0.45\linewidth}
    \center{\includegraphics[width=1\linewidth]{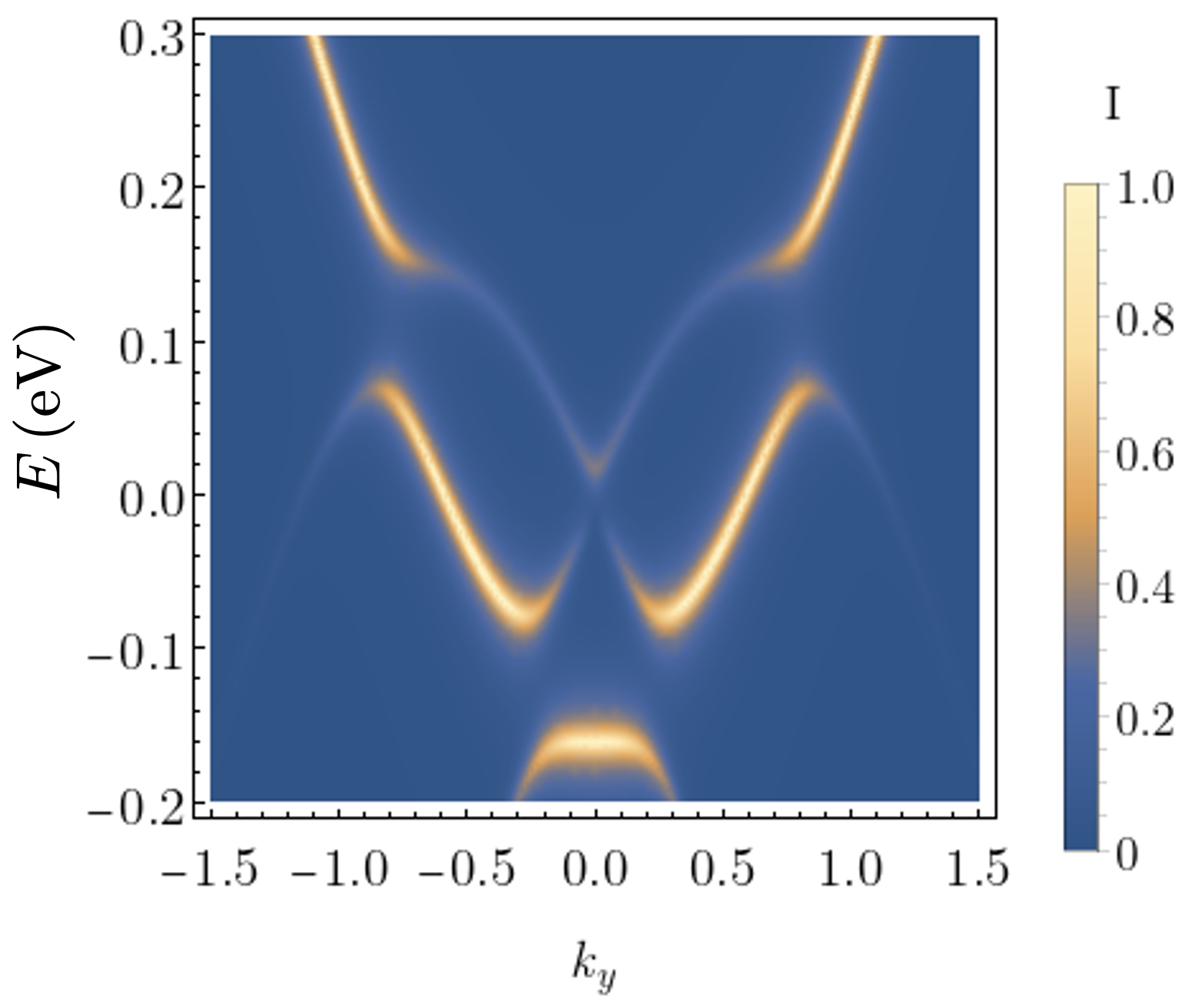}}
     \\b) ARPES spectra in the nodal region region $k_x=-2$. 
    \end{minipage} 
\caption{Band structure and ARPES spectrum of the PDW model in the nodal region with $\mathbf{Q}=\{ 0,0.6 \} $, $\Delta_\mathbf{Q}(k_x=-2)=0.05$, and the broadening for ARPES is chosen as $\Gamma=0.01$. The tight-binding parameters are same as in Fig.~\ref{fig:brillouin_zone}.}
\label{fig:pdw}
\end{figure}

Another possible model of pseudogap is the PDW state, which unlike its weak coupling version (the FFLO states) has been argued to be present at zero magnetic field \cite{Lee2014,AgterbergARCMP}. 
In the model of the PDW phase in Ref.~\cite{Lee2014}, the band structure is obtained by diagonalizing the following Hamiltonian
\begin{equation}
H(\mathbf{k})=\left(
    \begin{array}{ccc}
         \epsilon_c(\mathbf{k})& \Delta_\mathbf{Q} & \Delta_\mathbf{-Q}  \\
         \Delta_\mathbf{Q} &  -\epsilon_c(\mathbf{k+Q}) &0\\
          \Delta_\mathbf{-Q} & 0  &-\epsilon_c(\mathbf{k-Q})\\
    \end{array}
\right)\, ,
\end{equation}
The resulting three electronic bands and spectral functions near the nodal point are shown in Fig.~\ref{fig:pdw}.
The band structure is more complex, and does not fully coincide with the unhybridized band (as it did in the paramagnon fractionalization model in Fig.~\ref{fig:He}(b,d) and the experiment in Fig.~\ref{fig:He}(f)). In particular, the bands exhibit a gap at negative energies, which has not been seen in ARPES experiments. 

\bibliography{bibliography}

\end{document}